\documentclass{article}

\usepackage{arxiv}

\makeatletter
\renewcommand{\normalsize}{\@setfontsize\normalsize\@xpt{12}}
\makeatother

\usepackage[utf8]{inputenc}
\usepackage[T1]{fontenc}
\usepackage{url}
\usepackage{booktabs}
\usepackage{amsfonts}
\usepackage{amsmath}
\usepackage{nicefrac}
\usepackage{microtype}
\usepackage{graphicx}
\usepackage[super,sort&compress]{natbib}
\usepackage{doi}
\usepackage{enumitem}
\setlist{nosep}

\usepackage{siunitx}
\usepackage{xfp}
\usepackage{cleveref}

\usepackage{longtable}
\usepackage{array}
\usepackage{multirow}
\usepackage{threeparttable}
\usepackage{threeparttablex}
\usepackage{makecell}
\usepackage{pdflscape}
\usepackage[labelfont=bf]{caption}
\usepackage{float}

\usepackage{hyperref}
\usepackage{afterpage}
\usepackage{placeins}

\AtBeginDocument{%
  \abovedisplayskip=10pt plus 3pt minus 3pt
  \belowdisplayskip=10pt plus 3pt minus 3pt
  \abovedisplayshortskip=3pt plus 3pt
  \belowdisplayshortskip=6pt plus 3pt minus 2pt
}

\let\origlandscape\landscape
\renewcommand{\landscape}{%
  \origlandscape
  \renewcommand\TPTminimum{\textwidth}%
}


\newcommand{\bcsEligible}{16638}
\newcommand{\bcsM}{94}

\newcommand{\simLargeSamplePoisExposureMean}{1.91}
\newcommand{\simLargeSamplePoisExposureSd}{1.49}
\newcommand{\simLargeSamplePoisExposureQLow}{1}
\newcommand{\simLargeSamplePoisExposureQHigh}{3}
\newcommand{\simLargeSamplePoisExposureMed}{2.00}

\newcommand{\simLargeSamplePoisRrUnadjusted}{1.17}

\newcommand{\simLargeSampleNbExposureMean}{1.81}
\newcommand{\simLargeSampleNbExposureSd}{2.04}
\newcommand{\simLargeSampleNbExposureQLow}{0}
\newcommand{\simLargeSampleNbExposureQHigh}{3}
\newcommand{\simLargeSampleNbExposureMed}{1.00}
\newcommand{\simLargeSampleNbTheta}{1.30}

\newcommand{\simLargeSampleNbRejectionPerc}{0.9}

\newcommand{\simLargeSampleNbOutcomePerc}{4.5}
\newcommand{\simLargeSampleNbRrUnadjusted}{1.13}

\newcommand{\bcsResrawCBPSestimate}{1.53}

\newcommand{\bcsResrawnpCBPSconfLow}{1.08}
\newcommand{\bcsResrawnpCBPSconfHigh}{2.10}

\newcommand{\bcsResrawadjustedestimate}{1.50}

\newcommand{\bcsResrawunadjustedestimate}{1.79}

\newcommand{\bcsResrawunadjustedconfLow}{1.70}
\newcommand{\bcsResrawunadjustedconfHigh}{1.88}

\newcommand{\simCcPerfNegBinNpCbpsRelBiasRaw}{-15.2}

\newcommand{\simCcPerfNegBinNpCbpsRelBiasWin}{3.5}

\newcommand{\simCcPerfPoissonMultinomialRelBiasWin}{8.4}

\newcommand{\simCcPerfPoissonNpCbpsRelBiasRaw}{-26.1}

\newcommand{\simCcPerfPoissonNpCbpsRelBiasWin}{19.7}

\newcommand{\simCcPerfPoissonEnergyRelBiasWin}{6.3}

\newcommand{\simCcBalNegBinMultinomialRawMeanEss}{4667}

\newcommand{\simCcBalNegBinMultinomialRawFivePercEss}{4558}

\newcommand{\simCcBalNegBinCbpsRawMeanEss}{4679}

\newcommand{\simCcBalNegBinCbpsRawFivePercEss}{4605}

\newcommand{\simCcBalNegBinNpCbpsRawMeanEss}{3243}

\newcommand{\simCcBalNegBinGbmRawMeanEss}{4623}

\newcommand{\simCcBalNegBinGbmRawFivePercEss}{4511}

\newcommand{\simCcBalNegBinEnergyRawMeanEss}{3639}

\newcommand{\simCcBalPoissonMultinomialRawNineFiveRhoW}{0.026}
\newcommand{\simCcBalPoissonMultinomialRawMaxW}{0.122356093993802}

\newcommand{\simCcBalPoissonNpCbpsRawMeanEss}{1675}

\newcommand{\simCcBalPoissonNpCbpsRawFivePercEss}{453}

\newcommand{\compSpeedRelEnergy}{1285.3}
\newcommand{\compSpeedRelGBM}{273.0}
\newcommand{\compSpeedRelMultinomial}{3.9}
\newcommand{\compSpeedRelnpCBPS}{64.0}

\newcommand{\compSpeedTimeCBPS}{0.05}

\newcommand{\simCcBalNegBinCbpsGbmMultiMinEss}{%
  \fpeval{min(\simCcBalNegBinCbpsRawMeanEss,
              \simCcBalNegBinGbmRawMeanEss,
              \simCcBalNegBinMultinomialRawMeanEss)}%
}

\newcommand{\simCcBalNegBinCbpsGbmMultiMaxEss}{%
  \fpeval{max(\simCcBalNegBinCbpsRawMeanEss,
              \simCcBalNegBinGbmRawMeanEss,
              \simCcBalNegBinMultinomialRawMeanEss)}%
}

\newcommand{\simCcBalNegBinCbpsGbmMultiFivePercEssMin}{%
  \fpeval{min(\simCcBalNegBinCbpsRawFivePercEss,
              \simCcBalNegBinGbmRawFivePercEss,
              \simCcBalNegBinMultinomialRawFivePercEss)}%
}

\title{Inverse Probability Weighting of Count Exposures in the Presence of Missing Data: A Simulation Study}

\author{
  Martin N.~Danka\textsuperscript{1,2,*}
  \And
  Jessica K.~Bone\textsuperscript{2}
  \And
  George B.~Ploubidis\textsuperscript{1}
  \And
  Richard J.~Silverwood\textsuperscript{1}
}

\renewcommand{\shorttitle}{IPTW of Count Exposures in the Presence of Missing Data}

\hypersetup{
  pdftitle={Inverse Probability Weighting of Count Exposures in the Presence of Missing Data: A Simulation Study},
  pdfsubject={stat.ME},
  pdfauthor={Martin N. Danka, Jessica K. Bone, George B. Ploubidis, Richard J. Silverwood},
  pdfkeywords={iptw, propensity score, count data, multiple imputation, confounding},
  colorlinks=true,
  allcolors=black,
}

\begin{document}
\raggedbottom
\maketitle

\vspace{-0.5em}
\begin{center}
{\small
\textsuperscript{1}Centre for Longitudinal Studies, UCL Social Research Institute, University College London\\
\textsuperscript{2}Research Department of Behavioural Science and Health, Institute of Epidemiology and Health Care, University College London\\[6pt]
\textsuperscript{*}Corresponding author: \href{mailto:martin.danka.21@ucl.ac.uk}{martin.danka.21@ucl.ac.uk}
\enspace\href{https://orcid.org/0000-0003-0302-238X}{\includegraphics[scale=0.06]{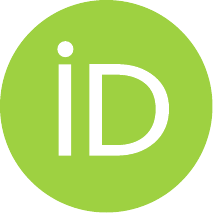}\,0000-0003-0302-238X}
}
\end{center}
\vspace{1.5em}

\begin{abstract}
Inverse probability of treatment weighting (IPTW) is widely used to estimate causal effects, but guidance is limited for count exposures. It is also unclear how IPTW performs when combined with multiple imputation in this context. In this study, we evaluated five IPTW methods applied to count exposures: multinomial binning, parametric and non-parametric covariate balancing propensity scores (CBPS, npCBPS), generalised boosted models (GBM), and energy balancing. Our simulations were informed by an example using data from the 1970 British Cohort Study, aiming to estimate the effect of psychological distress, measured as a count of symptoms at age 34, on self-reported longstanding illness at age 42. We compared these approaches on bias, coverage, effective sample size, and other metrics under truncated negative binomial and Poisson exposure distributions. We also assessed the performance of Rubin's rules under different missingness mechanisms. Under complete data, multinomial, CBPS, GBM, and energy weights produced low bias and near-nominal coverage, whereas npCBPS resulted in bias and poor coverage due to extreme weights. When data were missing completely at random, similar performance patterns were observed for IPTW with multiple imputation. Under missing at random, bias increased with higher missingness, but this was present for both IPTW and covariate-adjusted regression, possibly reflecting a limitation of the imputation model rather than a failure of IPTW. Overall, these findings support the use of multinomial, CBPS, GBMs, and energy weights for count exposures in similar settings while highlighting trade-offs between these methods and the need for imputation models accommodating right-truncated overdispersed counts.
\end{abstract}

\vspace{0.5em}
\keywords{iptw \and propensity score \and count data \and multiple imputation \and confounding}

\newpage

\section{Background}\label{sec1}

Observational datasets are a valuable resource for addressing causal questions related to health, with the potential to generate evidence from real-world settings. Yet, answering these questions is more challenging in the absence of randomisation, because associations between the exposure and the outcome may lack causal interpretation due to confounding. Confounders are common causes of the exposure and the outcome, and when they are known and measured, statistical approaches can be used to reduce confounding bias.

To reduce confounding due to measured variables, analysts typically rely on a working model of the data-generating process. Regression adjustment is a widely used approach, in which a regression model expresses the outcome as a function of the exposure and (potential) confounders. Valid adjustment therefore depends on how well this outcome model is specified. However, in some settings the analyst may be more willing to specify how confounders relate to the exposure mechanism, for example when outcomes are rare or otherwise challenging to model, or in outcome-wide analyses. \cite{vanderweele_outcome-wide_2020} In such cases, propensity score methods offer a compelling alternative to regression adjustment. The propensity score represents the probability of receiving an exposure level of interest given a person's covariate profile. Methods based on the propensity score aim to balance measured confounders across the exposure distribution, thereby approximating the balancing property of randomisation.\cite{rosenbaum_central_1983}

Building on the propensity score, inverse probability of treatment weighting (IPTW) is a popular method for estimating causal effects under confounding. This approach assigns each individual a weight inversely proportional to their propensity score, which aims to approximate a pseudopopulation where the exposure is independent of the observed confounders. The effect of interest can be estimated by implementing these weights through parametric approaches, such as weighted regression, or, less commonly, through non-parametric methods.\cite{robins_marginal_2000}

While IPTW is often applied to binary exposures, the procedure may be generalised to accommodate a range of distributions, including multicategorical and numeric exposures.\cite{robins_marginal_2000, gelman_propensity_2004} However, the estimation of weights for numeric exposures may be complex due to their diverse distributional characteristics, making correct exposure model specification more challenging. For this reason, weight estimation approaches that relax or avoid strict exposure model assumptions are particularly appealing in this setting. These include approaches based on binning the exposure,\cite{naimi_constructing_2014} directly optimising covariate balance,\cite{fong_covariate_2018, imai_covariate_2014} using machine learning to flexibly model the exposure-covariate relationships,\cite{zhu_boosting_2015} and, more recently, minimising broader measures of dependence between exposure and covariates.\cite{szekely_testing_2004, szekely_energy_2013, huling_independence_2024}

Despite these advancements, there is a notable gap in the application of IPTW to count exposures, which are prevalent in real-world settings. Count exposures, such as the number of hospital visits, medication doses, or symptom occurrences, are numeric exposures that take discrete, non-negative integer values, and, in many applications, are observed over only a small set of values. Their distributions may also be overdispersed, zero-inflated, or truncated. The distributional characteristics of counts may be challenging to accommodate in the IPTW framework, and approaches that avoid strict modelling assumptions are a compelling option. However, although some of these methods have been studied for similar distributions,\cite{naimi_constructing_2014, sack_inverse_2023} they were neither developed nor tested specifically for count exposures, and their appropriateness remains unclear in this context.

A further practical challenge for implementing IPTW is the pervasiveness of missing data in real-world datasets. When the probability of missingness given the observed data does not depend on any unobserved variables, the missing data mechanism is classified as missing at random (MAR). Under MAR, multiple imputation (MI) is a popular approach to mitigate bias that may arise when analyses are restricted to complete cases. IPTW can be incorporated in the MI framework using the `within-imputation' approach, where weights and effects are estimated separately within each imputed dataset, followed by pooling of the point estimates. \cite{qu_propensity_2009, seaman_inverse_2014, leyrat_propensity_2019, nguyen_multiple_2024}. However, simulation evidence largely comes from studies of binary exposures, so it remains unclear how well this strategy extends to count exposures. Additionally, little is known about the behaviour of flexible weight-construction approaches within this framework, thus limiting their applicability in real-world datasets.

To address these gaps, this study aims to investigate the performance of IPTW for count exposures with binary outcomes, evaluating five different approaches for weight estimation: a multinomial binning approach, parametric and non-parametric covariate balancing propensity scores (CBPS, npCBPS), generalised boosted models (GBM), and energy balancing. The second aim is to evaluate how these IPTW methods perform under missing data conditions when combined with MI using the `within-imputation' approach. This simulation study adheres to the ADEMP[I] framework.\cite{morris_using_2019} Our simulations were informed by a real-world example, estimating the effect of psychological distress symptom counts on the risk of self-reported longstanding illness in midlife.

The article is structured as follows. We begin by introducing IPTW notation and describing the five weight construction approaches considered in this study. We then describe the motivating example that informed the simulations. Next, we outline the simulation study design and report simulation results. We then return to the motivating cohort data to illustrate how the methods can be applied in practice. Finally, we interpret the findings and discuss implications and limitations.

\section{Inverse Probability of Treatment Weighting}\label{sec:iptw}

Central to IPTW is the concept of the propensity score, which we introduce before discussing IPTW and the weight estimation methods. Our aim is not to provide a detailed description of each method, but rather to give intuition, notation, and references for readers interested in further details. Throughout the manuscript, we assume that the exposure is not randomised, and we use the terms `exposure' and `treatment' interchangeably.

\subsection{The propensity score}

Let $A$ represent a binary exposure, taking values $a \in \{0, 1\}$. For each $a$, let $Y(a)$ denote the potential outcome that would be observed if, possibly counter to fact, $A = a$. $\boldsymbol{C}$ stands for a vector of baseline covariates, and we assume $\boldsymbol{C}$ satisfies the assumption of conditional exchangeability, $Y(a) \perp A \mid \boldsymbol{C}$. Then, the propensity score $e(\boldsymbol{c})$ is the conditional probability of being exposed given the covariate profile\cite{rosenbaum_central_1983}:
\[
e(\boldsymbol{c}) = \Pr(A = 1 \mid \boldsymbol{C} = \boldsymbol{c})
\]
This propensity score can be estimated using both non-parametric and parametric approaches. A common approach is to model the binary exposure as a function of the covariates using logistic regression, where the propensity score is estimated by the predicted probabilities. The estimated propensity score can be used to adjust for differences in covariate distributions between treatment groups, thereby mimicking the conditions of a randomised experiment. This can be achieved through stratification on the score, matching individuals with similar scores, or weighting, which is the focus of this paper.

The propensity score has later been generalised to other exposure distributions. \cite{gelman_propensity_2004, imbens_role_2000} To describe this, we now allow $A$ to denote a generic exposure taking values $a$ in a set $\Psi$ (which may be discrete or continuous). The generalised propensity score is defined as the conditional probability mass or density function of $A$ given the covariates:
\[
r(a, \boldsymbol{c}) = f_{A \mid \boldsymbol{C}}(a \mid \boldsymbol{c}), \quad a \in \Psi,
\]
where $f_{A \mid C}$ denotes the conditional probability mass function (for discrete $A$) or probability density
function (for continuous $A$) of $A$ given $\boldsymbol{C} = \boldsymbol{c}$. In the binary case with $A \in \{0,1\}$, the standard propensity score is recovered as $e(\boldsymbol{c}) = r(1, \boldsymbol{c})$.

\subsection{Exposure weights}

We now give an overview of IPTW, following the formulation of Robins, Hern\'{a}n, and Brumback.\cite{robins_marginal_2000} IPTW assigns each individual a weight inversely proportional to the propensity score. For binary exposures, the weights are determined as follows:
\[
w =
    \begin{cases}
        \dfrac{1}{e(\boldsymbol{c})} & \text{if } A = 1,\\[10pt]
        \dfrac{1}{1 - e(\boldsymbol{c})} & \text{if } A = 0.
    \end{cases}
\]
Assigning weights in this manner ensures that, given each person's covariate profile, individuals underrepresented in their exposure group are upweighted, whereas overrepresented individuals are downweighted. Re-weighting in this way approximates a pseudopopulation where $A \perp \boldsymbol{C}$, thus confounding is removed.

More generally, the weights can also be derived from the generalised propensity score:
\[
w = \dfrac{1}{r(a, \boldsymbol{c})} = \dfrac{1}{f_{A \mid \boldsymbol{C}}(a \mid \boldsymbol{c})}
\]
In practice, the inverse probability weights can be unstable if some individuals have very small estimated
propensity scores. A common modification is to use stabilised weights, which replace the numerator with the marginal
probability (or density) of the observed exposure, denoted $f_A(a)$\cite{cole_constructing_2008}:
\[
w_s =
\dfrac{f_A(a)}{f_{A \mid \boldsymbol{C}}(a \mid \boldsymbol{c})},
\]
Under the identifying assumptions of conditional exchangeability, positivity, consistency, and no interference,\cite{goetghebeur_formulating_2020} these exposure weights can be used to estimate average marginal causal effects.

For a binary exposure, the average treatment effect ($ATE$), is defined as $ATE = E\{Y(1)\} - E\{Y(0)\}$. The non-parametric IPTW estimator for the $ATE$ replaces mean potential
outcomes with weighted averages of the observed outcomes for the exposed and unexposed:
\[
\widehat{\text{ATE}}_{\text{IPTW}}
= \widehat{E}\{Y(1)\} - \widehat{E}\{Y(0)\}
= \frac{\sum_{i=1}^n w_i Y_i A_i}{\sum_{i=1}^n w_i A_i}
  - \frac{\sum_{i=1}^n w_i Y_i (1 - A_i)}{\sum_{i=1}^n w_i (1 - A_i)} .
\]
For a non-binary exposure (including counts), IPTW can target the average (marginal) dose-response function $\mu(a) = E\{Y(a)\}$,\cite{gelman_propensity_2004} and this can be used to define average causal effects as contrasts $\mu(a) - \mu(a')$ between two exposure levels $a$ and $a'$. The dose-response can be expressed using a marginal structural model,\cite{robins_marginal_2000}
\[
\mu(a) = E\{Y(a)\} = m(a; \boldsymbol{\delta}),
\]
where $m(\cdot)$ is a chosen working model (for example, linear, spline-based, or categorical in $a$).
The parameters $\boldsymbol{\delta}$ can then be estimated using IPTW, typically by fitting an outcome model re-weighted by the inverse probability weights. If the marginal structural model is linear, $E\{Y(a)\} = \delta_0 + \delta_1 a$, the dose-response has a constant slope, so that
$\delta_1$ represents a constant average change in $\mu(a)$ per one-unit increase in $A$. Other causal effects, such as the marginal risk or odds ratio, can be expressed in a similar manner. \cite{colnet_risk_2025}

In practice, IPTW can be challenging to implement for numeric exposures due to its sensitivity to model misspecification, as both the conditional and marginal mass or density functions for stabilised weights must be correctly specified, and extreme weights are a common issue even after stabilisation. While this issue is not specific to counts, it is amplified by the challenges of specifying discrete distributions that accommodate overdispersion, truncation, zero inflation, and other features. Several alternatives relax this parametric assumption, which we explore further.

\subsection{Binning approaches}

Multicategorical exposures can be modelled directly using multinomial or ordinal regression, which allows the
derivation of stabilised exposure weights. This approach can be extended to numeric exposures by first
grouping $A$ into $K$ categories (e.g. using quantiles or theory-driven cut--points) and then modelling the resulting
multicategory exposure.\cite{naimi_constructing_2014} Using multinomial regression, the denominator of the stabilised weight for an
individual with exposure level $A = k$ and covariates $\boldsymbol{C} = \boldsymbol{c}$ is the estimated
conditional probability
\[
\hat f_{A \mid \boldsymbol{C}}(k \mid \boldsymbol{c})
= \Pr(A = k \mid \boldsymbol{C} = \boldsymbol{c})
= \frac{\exp\{\hat\beta_{0k} + \boldsymbol{c}^\top \hat{\boldsymbol{\beta}}_{k}\}}
       {\sum_{j=1}^K \exp\{\hat\beta_{0j} + \boldsymbol{c}^\top \hat{\boldsymbol{\beta}}_{j}\}} .
\]
The numerator of the stabilised weight is the estimated marginal probability of receiving exposure level $k$,
$\hat f_A(k) = \widehat{\Pr}(A = k)$.

\subsection{Covariate balancing approaches}

The propensity score has a dual role: it represents the conditional probability of receiving a given exposure level given covariates, and, under correct model specification, it also acts as a balancing score, meaning the resulting weights balance the distribution of covariates across exposure levels.\cite{rosenbaum_central_1983} The covariate balancing propensity score (CBPS) methods exploit this dual role, attempting to find propensity scores or weights optimised on this balancing property.\cite{fong_covariate_2018, imai_covariate_2014}

For count exposures, we consider the extension by Fong et al.,\cite{fong_covariate_2018} originally proposed for continuous treatments. The exposure $A$ is first
standardised to $A_i^{*}$, and covariates are centred and orthogonalised to give
$\boldsymbol{C}_i^{*}$, so the stabilised weight for each individual takes the form:
\[
w_i = \frac{f_{A^{*}}(A_i^{*})}
                 {f_{A^{*} \mid \boldsymbol{C}^{*}}(A_i^{*} \mid \boldsymbol{C}_i^{*})}
\]
CBPS then uses constrained optimisation to identify weights $w_i$ that satisfy a
balancing constraint:
\[
\sum_{i=1}^N w_i A_i^{*} \boldsymbol{C}_i^{*} = \boldsymbol{0},
\]
so that, in the weighted sample, the standardised exposure and covariates are uncorrelated.

This generalisation of the CBPS can be employed parametrically or non-parametrically. In the parametric version, the marginal density $f_{A^{*}}$ is assumed to be standard normal, and $f_{A^{*} \mid \boldsymbol{C}^{*}}$ is a conditional normal density obtained from a linear regression of $A^{*}$ on $\boldsymbol{C}^{*}$. The regression parameters are estimated by solving a system of equations that combines (i) a standard normal--regression condition for the residual variance with (ii) the covariate balancing condition above, which can be viewed as an application of the generalised method of moments. Because estimation explicitly targets balance rather than only model fit, we expect this approach to retain some robustness to misspecification of the parametric model and marginal distribution for $A$, making the approach potentially applicable to count exposures provided that balance is achieved.

The non-parametric implementation (npCBPS) does not require specifying the
distributional form of $f_{A^{*}}$ or $f_{A^{*} \mid \boldsymbol{C}^{*}}$. Instead,
it uses constrained optimisation to directly find positive weights $w_i$ that meet
the balancing condition above and also satisfy additional constraints capturing
desirable properties of stabilised weights. In the weighted sample, the
standardised exposure and covariates retain their marginal means (that is, zero), and the total weight
equals $N$:
\[
\sum_{i=1}^N w_i \boldsymbol{C}_i^{*} = \boldsymbol{0}, \quad
\sum_{i=1}^N w_i A_i^{*} = 0, \quad
\sum_{i=1}^N w_i = N, \quad
w_i > 0 \ \text{for all } i.
\]
Among the weight vectors satisfying these constraints, npCBPS uses an empirical likelihood criterion to identify weights that are, as far as the constraints allow, close to equal (i.e. each $w_i$ close to 1), thereby attempting to avoid overly extreme weights.

\subsection{Generalised boosted models}

Generalised boosted models (GBMs) use the gradient boosting algorithm to flexibly estimate regression functions. \cite{friedman_greedy_2001} They can be used to estimate the exposure--confounder relationships, and thus estimate the propensity score and exposure weights. \cite{mccaffrey_propensity_2004, mccaffrey_tutorial_2013} We consider the extension that was originally proposed for continuous exposures by Zhu et al. \cite{zhu_boosting_2015} In this context, a commonly used working model for the generalised propensity score assumes a linear mean function
\[
A_i = \boldsymbol{C}_i^\top \boldsymbol{\beta} + \varepsilon_i,
\quad \varepsilon_i \sim \mathcal{N}(0,\sigma^2),
\]
so that the conditional density $f_{A \mid \boldsymbol{C}}(A_i \mid \boldsymbol{C}_i)$
can be obtained from a normal model, and the marginal density $f_A(A_i)$ can be
specified parametrically or estimated by kernel density estimation.

GBMs replace the linear predictor with a flexible, non-parametric regression function
\[
A_i = m(\boldsymbol{C}_i) + \varepsilon_i,
\]
where $m(\boldsymbol{C}_i)$ is represented as a sum of simple regression trees. Each
tree partitions the covariate space into regions $R_{bj}$ and assigns a constant
prediction $c_{bj}$ within each region, so that
\[
m(\boldsymbol{C}_i)
= m_0 + \sum_{b=1}^{M} \sum_{j=1}^{K_b} c_{bj}\, I\{\boldsymbol{C}_i \in R_{bj}\},
\]
with $m_0$ the initial mean of $A$, $M$ the total number of trees, $K_b$ the number
of terminal nodes (regions) in tree $b$, and $I\{\boldsymbol{C}_i \in R_{bj}\}$ an
indicator that observation $i$ falls in region $R_{bj}$. The algorithm starts from
$m_0$ and then adds trees sequentially, each fit to the current residuals
(pseudo-residuals), so that each $c_{bj}$ acts as a small correction to the current
fit in region $R_{bj}$.

The number of trees $M$ controls the complexity of the model. In standard
applications, $M$ can be chosen by cross-validation. For propensity score estimation,
GBMs are instead tuned to optimise covariate balance in the weighted sample. For
binary exposures, one can stop boosting at the iteration that minimises the average
standardised absolute mean difference in covariates between exposure groups. For
continuous exposures, $M$ can be chosen to minimise the average absolute weighted correlation between the
exposure and the covariates.

\subsection{Energy balancing}

While the previously discussed methods either relax or are robust to certain restrictive parametric assumptions, they nonetheless require investigators to choose which moments of the covariates and exposure should be decorrelated, and there are no established guidelines for making this choice. Energy balancing relies on energy distance-based metrics, enabling researchers to circumvent the need to pre-specify particular moments for decorrelation. Intuitively, energy distance quantifies how (dis)similar two distributions are by comparing average pairwise distances within observations drawn from the same distribution with those between observations from different distributions. It is zero when the two distributions
are identical and increases as they diverge, capturing differences not only in means but also in higher-order moments. \cite{szekely_energy_2013}

Building on this metric, Huling and Mak \cite{huling_energy_2024} proposed energy balancing for binary
exposures, which uses constrained optimisation to find weights that minimise the energy distance between the weighted
covariate distribution in each exposure group and the covariate distribution in the overall
sample. This idea was extended to continuous exposures, defining
independence weights (also called distance covariance optimal weights).\cite{huling_independence_2024} In this setting,
the weights are obtained by minimising a criterion that combines (i) a weighted distance covariance
term $V_{n,\boldsymbol{w}}^2(\boldsymbol{C}, A)$, which is equal to zero if the exposure $A$ and covariates $\boldsymbol{C}$ are
independent under the weights, and (ii) two energy distance terms $\mathcal{E}\{F_{\boldsymbol{C},\boldsymbol{w}}^n, F_{\boldsymbol{C}}^n\}$ and $\mathcal{E}\{F_{A,\boldsymbol{w}}^n, F_A^n\}$ that penalise departures of
the weighted marginal distributions of $A$ and $\boldsymbol{C}$ from their original empirical
marginals. Formally, the weights $\boldsymbol{w}$ minimise a criterion of the form:
\[
D(\boldsymbol{w})
= V_{n,\boldsymbol{w}}^2(\boldsymbol{C}, A)
  + \mathcal{E}\{F_{\boldsymbol{C},\boldsymbol{w}}^n, F_{\boldsymbol{C}}^n\}
  + \mathcal{E}\{F_{A,\boldsymbol{w}}^n, F_A^n\},
\]
The optimisation is carried out
subject to $w_i \ge 0$ and $\sum_{i=1}^N w_i = N$, so that the average weight is 1. As with
CBPS and npCBPS, these constraints yield stabilised weights by construction.

\section{Motivating example}\label{sec2}
We now introduce the motivating example, which reflects some of the methodological challenges faced when applying IPTW to counts. The data come from the 1970 British Cohort Study (BCS70). BCS70 is a multidisciplinary birth cohort of over 17{,}000 individuals born in a single week in 1970 in England, Scotland, and Wales, with 11 completed sweeps of data collection to date. Further details are available in the cohort profiles \cite{elliott_cohort_2006, sullivan_cohort_2023} and on the Centre for Longitudinal Studies website (\url{https://cls.ucl.ac.uk/cls-studies/1970-british-cohort-study/}).

Across multiple data collection sweeps, participants completed the Malaise Inventory, a measure of psychological distress. A nine-item version was introduced at age 34 (2004 sweep), with binary (`yes'/`no') responses to symptoms such as worrying often, feeling miserable or depressed, and feeling tired most of the time, yielding a symptom count ranging from 0 to 9. Previous work has demonstrated scalar invariance of this shortened measure across time, cohorts, and gender in UK birth cohorts. \cite{mcelroy_harmonisation_2020, ploubidis_longitudinal_2019} At age 42 (2012 sweep), participants were asked a series of questions about their health. For simplicity in this motivating example, we focus on a single broad binary indicator of having a longstanding illness. This was marked as present if a participant answered `yes' to having ``any physical or mental health conditions or illnesses lasting or expected to last 12 months or more".

Our aim is to estimate the effect of psychological distress at age 34 on the presence of longstanding illness at age 42 (2012 sweep). This relationship is likely confounded by many influences, with many potential confounders available in the dataset, as outlined in Table~\ref{tab:bcs70_measures} in the Supplementary Information. These include prior health and health behaviours (for example, past longstanding illness, physical activity), adult socioeconomic situation (occupation and income), sociodemographic characteristics (sex at birth, partnership), and early-life and developmental influences (including parental social class, overcrowding, housing tenure, parental education and income, birthweight, maternal smoking in pregnancy, childhood cognition, and adolescent psychological distress). The estimand is the causal risk ratio (RR), and we wish to estimate this using IPTW. We focus on the target population of individuals born in 1970 in England, Scotland, or Wales, who remained alive and did not emigrate by age 42. This problem is further used to inform the simulation design.

\section{Simulation study design}\label{sec:simdesign}

\subsection{Data-generating mechanisms}

To evaluate the performance of the five IPTW approaches with count exposures, we began by generating 2,000 datasets, each having 5,000 observations. The number of repetitions was determined to give about 0.005 Monte Carlo standard error (SE) on coverage. \cite{morris_using_2019}

Three covariates $C_{1}$, $C_{2}$ and $C_{3}$ were sampled from a multivariate normal distribution $\mathbf{C} \sim \mathcal{N}(\mathbf{0}, \boldsymbol{\Sigma})$, where the diagonal elements (the variances) were set to $\Sigma_{ii} = 1$ and the off-diagonal elements (corresponding to pairwise correlations) to $\Sigma_{ij} = 0.3$ for all $i \neq j$. After sampling, $C_{1}$ was dichotomised by setting it to $0$ where $C_{1} \leq 0$ and to $1$ where $C_{1} > 0$, to keep the proportions approximately equal to $0.5$, resulting in a variable resembling sex at birth in cohort datasets. $C_{2}$ and $C_{3}$ were retained as continuous variables.

Exposure $A$ was generated under two data-generating mechanisms (DGMs). The first mechanism aimed to produce a truncated conditional negative binomial distribution, motivated by the Malaise Inventory, using minimal adjustments to avoid overreliance on the specific characteristics of the motivating example. The second mechanism focused on a truncated conditional Poisson distribution with similar properties but no overdispersion. The expected count for each observation was given by
\[
\lambda_i = \exp(\beta_0 + \beta_1 C_{i,1} + \beta_2 C_{i,2} + \beta_3 C_{i,3})
\]
where $\beta_0=\ln{1.5}$ was chosen pragmatically to generate a marginal exposure distribution broadly similar to that observed for the Malaise Inventory at age 34 in BCS70 (mean count $1.7$), and $\beta_1=0.4$, $\beta_2=0.1$, and $\beta_3=0.1$ to keep most counts within close ranges. Under the first DGM, the exposure variable was then sampled from a negative binomial distribution $A \sim NegBin(\lambda_i,\ \ k)$, with the dispersion parameter set to the observed marginal $\hat{k} = 1.3$ of the Malaise Inventory at age 34 in BCS70. The second mechanism used the same model but simulated $A\sim Pois(\lambda_i)$. Because count distributions are often truncated in real-world datasets, we employed a rejection sampling procedure, resampling all $A_i>10$ under both mechanisms.

The exposure and covariates were then used to simulate a binary outcome $Y$. The conditional probability of $Y$ was expressed using a log-binomial model:
\[
\ln\pi_i = \gamma_0 + \gamma_1 A_{i} + \gamma_2 C_{i,1} + \gamma_3 C_{i,2} + \gamma_4 C_{i,3}
\]
Here, $\gamma_0 = \ln0.03$, setting the baseline probability to 3\% to roughly reflect the prevalence of
coronary heart disease in England in 2023. \cite{office_for_health_improvement__disparities_cardiovascular_2024}. The covariate effects were set to $\gamma_2 = \ln 1.4$
and $\gamma_3 = \gamma_4 = \ln 1.1$. The dose-response effect of interest was varied across the values $\gamma_1 \in {\{\ln 1.0, \ln 1.1,\ln 1.2}\}$. The covariate coefficients and the intercept were selected to keep $\pi_{i} < 1$ for all $i$, which was empirically verified for all simulated datasets. The outcome was then simulated as $Y_{i} \sim Bernoulli(\pi_{i})$.

\subsection{Missing data mechanisms}

To introduce missing data, we first simulated complete datasets and then induced missingness by amputation.\cite{oberman_toward_2024, morris_comment_2024} We aimed
to approximate a scenario commonly observed in birth cohorts, including BCS70, where
missingness tends to increase progressively with each sweep of data collection. We therefore
introduced missing data in the confounders $C_2$ and $C_3$, the exposure $A$, and the
outcome $Y$, while the binary confounder $C_1$ remained fully observed, resembling a
baseline variable such as sex at birth that is typically known for all cohort members. The
per-variable probability of missingness $p_i$ was specified to increase in the order
$p_{C_2} < p_{C_3} < p_{A} < p_{Y}$. We also aimed to control the overall level of
missingness, defined as the probability that an individual has at least one missing value,
denoted by $\phi$. We considered three target overall levels of missingness,
$\phi^\ast \in \{0.20, 0.40, 0.60\}$, across two scenarios, MCAR and MAR.

Under MCAR, the variables $C_2$, $C_3$, $A$, and $Y$ were independently set to missing with
variable-specific probabilities $p_i$, $i = 1,\dots,4$, ordered as above. The overall
missingness probability under MCAR is
\[
\phi_{\text{MCAR}} = 1 - \prod_{i=1}^4 (1 - p_i).
\]
We imposed a simple linear structure
\[
p_i = p_0 + (i-1)d,\quad i = 1,\dots,4,
\]
with $p_0 \ge 0$, $d \ge 0$ and $p_0 + 3d < 1$, so that the marginal proportion of
missingness increased across variables. To ensure we controlled both the per-variable $p_i$
and the overall $\phi$, we used constrained optimisation to select $(p_0, d)$ so that the
average realised $\bar\phi \approx \phi^\ast$, minimising $(\bar\phi - \phi^\ast)^2$.

Under MAR, we again allowed $C_2$, $C_3$, $A$, and $Y$ to be missing but now generated
missingness indicators from logistic models in which missingness depended on the remaining
variables. Let $M_{ij} = 1$ if variable $i$ is missing for individual $j$ and $0$ otherwise.
For simplicity, we omit the individual index $j$ and write $M_i$ for the missingness indicator of variable $i$, so that
$p_i = E(M_i)$. The missingness mechanisms were specified as

\[
\begin{aligned}
\text{logit}\{\Pr(M_{C_2} = 1 \mid C_1)\}
  &= \gamma_{0,C_2} + C_1, \\
\text{logit}\{\Pr(M_{C_3} = 1 \mid C_1, C_2)\}
  &= \gamma_{0,C_3} + C_1 + C_2, \\
\text{logit}\{\Pr(M_{A} = 1 \mid C_1, C_2, C_3)\}
  &= \gamma_{0,A} + C_1 + C_2 + C_3, \\
\text{logit}\{\Pr(M_{Y} = 1 \mid C_1, C_2, C_3, A)\}
  &= \gamma_{0,Y} + C_1 + C_2 + C_3 + A.
\end{aligned}
\]

The intercepts $\gamma_{0,i}$ control the marginal missingness probabilities
$p_i = E(M_i)$, but because the $M_{ij}$ are now dependent, the overall missingness rate
under MAR,
\[
\phi_{\text{MAR}} = 1 - E\Bigl\{\prod_{i=1}^{4} (1 - M_i)\Bigr\},
\]
is no longer equal to $1 - \prod_i (1 - p_i)$.

Therefore, to impose the same linear progression $p_i = p_0 + (i-1)d$ while maintaining the
target $\phi^\ast$, we used constrained optimisation with a Monte Carlo approximation to the
MAR objective: for a candidate $(p_0, d)$, we (i) solved for $\gamma_{0,i}$ that yielded the
requested $p_i$ and (ii) evaluated $\hat\phi_{\text{MAR}}$ empirically in a fixed large
simulated sample ($N = 1{,}000{,}000$). We then chose $(p_0, d)$ to make
$\hat\phi_{\text{MAR}} \approx \phi^\ast$. Full details are provided in
the Supplementary Information, with parameter values in Table~\ref{tab:missingness_parameters}.

\subsection{Effect and weight estimation}

The estimand of interest was the $\ln RR$ per unit increase in the exposure. This was estimated using a Poisson regression model for the outcome, using a sandwich estimator for variance (a procedure known as modified Poisson regression).\cite{zou_modified_2004} The model was reweighted using stabilised treatment weights. We compared five methods for constructing these weights: binning with weights constructed via a multinomial model (further referred to as `multinomial weights'), CBPS, npCBPS, GBM, and energy balancing (distance covariance optimal weights). For each weighting method, analyses were conducted using both the estimated stabilised treatment weights and a winsorised version of these weights, with winsorisation at the 99th percentile. To provide a comparison against a method with known properties, we also estimated the effect using unweighted regression with simple covariate adjustment. Lastly, we provide results for the unadjusted regression model, without any account for confounding, for further comparison.

The weighting methods were implemented using the {\tt WeightIt} R package.\cite{greifer_weightit_2024} The package implementations of the methods broadly match the descriptions in Section \ref{sec:iptw}, with a few practical details noted here. Prior to fitting the multinomial model to derive weights, we identified exposure levels with individual prevalence below 1\% in each simulated dataset. Whenever this occurred, the affected levels were confined to the extreme upper tail of the exposure distribution. These levels were then combined to avoid sparse multinomial categories. The exposure was only binned this way for constructing the multinomial weights, but the original count exposure entered the (re-weighted) model for effect estimation. In the {\tt WeightIt} implementation of CBPS, the marginal exposure mean and variance enter the moment conditions as parameters rather than being fixed to their observed values. For GBMs, the algorithm was similar to that described in Zhu et al.,\cite{zhu_boosting_2015} with two modifications to the balance-based stopping rule. First, the criterion was set to minimise the root-mean-squared absolute weighted correlation instead of the average absolute weighted correlation between the exposure and covariates, thus larger imbalances were penalised more strictly. Second, the default {\tt WeightIt} implementation computes this correlation in the full sample rather than using a bootstrap-based estimate.

For incomplete datasets, missing data were handled using MI by chained equations. All analysis variables were included in the imputation models. The continuous and count-type variables ($C_2$, $C_3$, $A$) were imputed using predictive mean matching, and the binary outcome $Y$ was imputed using logistic regression, as implemented in the {\tt mice} R package. \cite{van_buuren_mice_2011} In line with general recommendations, the number of imputations was chosen to equal the realised $\phi_r \times 100$ in each simulated dataset. \cite{white_multiple_2011} The imputed datasets were analysed using the `within-imputation' approach, combining point estimates and sandwich variances using Rubin's rules.\cite{rubin_multiple_2004} To reduce computational burden, analyses under missing data were only conducted for the DGMs with true RR = 1.1.

\subsection{Performance metrics}

The primary metrics of interest were
bias and coverage, although empirical and model-based SE were also provided, following definitions in Morris et al. \cite{morris_using_2019}

Weighting methods may produce large weights, with reduced precision being the potential consequence. Effective sample size (ESS) is a commonly used summary of this loss of precision, representing the size of a hypothetical unweighted sample that would provide the same precision as the weighted sample. This is obtained as $ESS =\frac{\left(\sum w\right)^2}{\sum w^2}$, where $w$ stands for the weights.\cite{kong_sequential_1994, mccaffrey_propensity_2004} In this study, we provided the average ESS for each method across the simulated datasets, its standard deviation, and the 5th and 95th percentiles.

To assess if the weights successfully decorrelated the treatment and the covariates, we provided metrics based on the absolute weighted treatment-covariate Pearson (or point-biserial) correlations, \cite{austin_assessing_2019} averaged across the three covariates in each dataset. Additionally, the dependence metric was used as a composite summary of the weights, providing a relative comparison of how well different methods reduced residual dependence between $A$ and $C$ while preserving the original marginal distributions. Lower values of the dependence metric indicate better-performing weights. We also reported the energy-distance components separately to assess deviations from the original marginal distributions of $A$ and $C$.\cite{huling_independence_2024} Because some methods optimise directly on these metrics, they were not used as evidence of superior performance for those methods.

The simulations were conducted in R v4.4 under the UCL Myriad High Performance Computing facility. We also assessed the computational speed of each weighting method locally, summarising the relative time required for weight estimation over 20 simulated datasets. Further details of the simulation implementation and the computation time assessment are provided in the Supplementary Information.

\section{Simulation results}\label{sec3}

\subsection{Simulated datasets}
\begin{figure}[t]
  \centering
  \includegraphics[width=\linewidth]{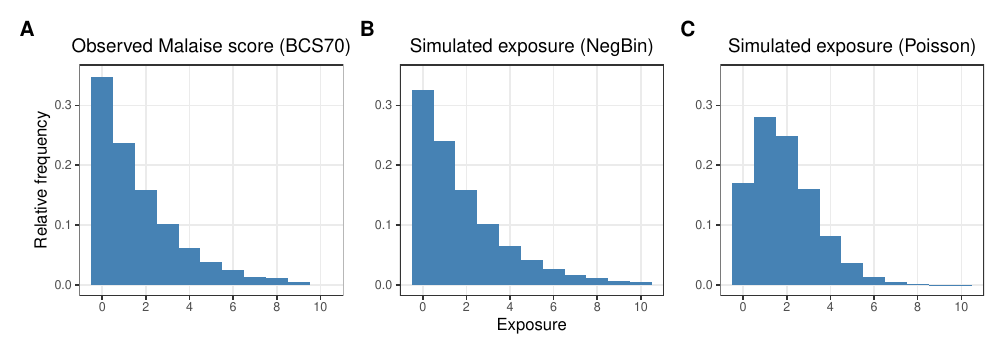}
  \caption{Histograms of the Malaise score and the simulated exposures.
    \textbf{A}, Observed Malaise score at age 34 in the 1970 British Cohort Study (N = 9 596; range 0--9).
    \textbf{B}, Simulated exposure (N = 1,000,000) from a conditional negative binomial distribution, truncated by resampling values greater than 10.
    \textbf{C}, Simulated exposure (N = 1,000,000) from a conditional Poisson distribution, truncated by resampling values greater than 10. BCS70 -- 1970 British Cohort Study; NegBin -- Negative Binomial.}
  \label{fig:exposure_comparison}
\end{figure}

To approximate and further characterise the distributions implied by each DGM, we generated a large sample of 1,000,000 observations. The achieved marginal distribution of the simulated truncated negative binomial exposure yielded the average count of
\simLargeSampleNbExposureMean\
(SD = \simLargeSampleNbExposureSd,
Mdn = \simLargeSampleNbExposureMed,
IQR from \simLargeSampleNbExposureQLow\ to \simLargeSampleNbExposureQHigh,
overdispersion parameter $k$ = \simLargeSampleNbTheta,
panel~B of Figure~\ref{fig:exposure_comparison}),
which was similar in location and scale to that of Malaise Inventory (M = 1.7, SD = 1.90, Mdn = 1.00, IQR 0 to 3, overdispersion parameter $k$ = 1.29, panel~A of Figure~\ref{fig:exposure_comparison}). The marginal truncated Poisson distribution had an average count of
$\lambda$ = \simLargeSamplePoisExposureMean\ (SD = \simLargeSamplePoisExposureSd,
Mdn = \simLargeSamplePoisExposureMed,
IQR \simLargeSamplePoisExposureQLow
to \simLargeSamplePoisExposureQHigh,
panel~C of Figure~\ref{fig:exposure_comparison}).
Most exposure values remained within the pre-specified range ($\le$ 10), with  only \simLargeSampleNbRejectionPerc{}\% values truncated via resampling for the negative binomial exposure and fewer than 0.01\% for the Poisson exposure. When the true RR was 1.10, the marginal prevalence of the simulated outcome was \simLargeSampleNbOutcomePerc{}\% for both the negative binomial and Poisson exposures. The confounded RR per unit increase in the exposure was \simLargeSampleNbRrUnadjusted{} under the negative binomial exposure and \simLargeSamplePoisRrUnadjusted{} under the Poisson exposure.

\subsection{Performance under complete data}
Our simulations produced comparable results for all metrics across the three effect sizes. Therefore, unless otherwise stated, we present results for the scenario in which the true unconfounded RR = 1.1. An interactive table of all results can be found at \url{martindanka.github.io/iptw-sim-shinylive/}.\,\textsuperscript{\dag}
\begingroup\renewcommand{\thefootnote}{\dag}%
\footnotetext{Please note that the application may require several minutes to load on first use.}%
\endgroup

Performance metrics under no missingness are shown in Table~\ref{tab:sim_cc_performance}. CBPS, energy weights, GBM, and multinomial weights all demonstrated good performance across the two exposure DGMs in terms of low bias (absolute value of relative bias below 3\%). In contrast, npCBPS underestimated the true effect, with relative bias amounting to \simCcPerfNegBinNpCbpsRelBiasRaw \% of the true effect for the negative binomial exposure and \simCcPerfPoissonNpCbpsRelBiasRaw\% for the Poisson exposure. The sandwich variance estimator also provided near-nominal coverage for all methods with low bias.

Winsorisation of weights at the 99\textsuperscript{th} percentile had relatively modest impact on performance. For multinomial, CBPS, GBM and energy weights, winsorisation led to slightly larger bias, particularly under the Poisson exposure DGM where relative bias ranged from \simCcPerfPoissonEnergyRelBiasWin\% to \simCcPerfPoissonMultinomialRelBiasWin\%. Coverage for these methods remained near the nominal value, with slightly smaller standard errors compared with raw weights. For npCBPS, winsorisation reduced the magnitude of bias and improved coverage, but performance remained worse than for the other weighting schemes (relative bias of \simCcPerfNegBinNpCbpsRelBiasWin\% for the negative binomial and \simCcPerfPoissonNpCbpsRelBiasWin\% for the poisson DGM).

\afterpage{
\begin{landscape}\begin{table}[htbp]
\scriptsize
\begin{threeparttable}
\caption{\label{tab:sim_cc_performance}Performance metrics for the simulations under complete data and RR = 1.1}
\begin{tabular}[t]{lcccccccccc}
\toprule
\multicolumn{1}{c}{ } & \multicolumn{5}{c}{\textbf{Raw weights}} & \multicolumn{5}{c}{\textbf{Winsorised weights (99\textsuperscript{th} percentile)}} \\
\cmidrule(l{3pt}r{3pt}){2-6} \cmidrule(l{3pt}r{3pt}){7-11}
\textbf{Method} & \textbf{Bias (MCSE)} & \textbf{Rel. Bias (MCSE)} & \textbf{Emp. SE} & \textbf{Model SE} & \textbf{Coverage (MCSE)} & \textbf{Bias (MCSE)} & \textbf{Rel. Bias (MCSE)} & \textbf{Emp. SE} & \textbf{Model SE} & \textbf{Coverage (MCSE)}\\
\midrule
\addlinespace[0.3em]
\multicolumn{11}{l}{\textbf{Negative binomial exposure}}\\
\hspace{1em}Unadjusted & 0.026 (0.001) & 27.1 (0.6) & 0.026 & 0.026 & 0.811 (0.009) & — & — & — & — & —\\
\hspace{1em}Adjusted & -0.001 (0.001) & -0.9 (0.6) & 0.028 & 0.027 & 0.953 (0.005) & — & — & — & — & —\\
\hspace{1em}Multinomial & -0.001 (0.001) & -1.0 (0.7) & 0.032 & 0.031 & 0.950 (0.005) & 0.003 (0.001) & 2.9 (0.7) & 0.030 & 0.029 & 0.951 (0.005)\\
\hspace{1em}CBPS & -0.002 (0.001) & -2.2 (0.7) & 0.032 & 0.031 & 0.951 (0.005) & 0.002 (0.001) & 2.0 (0.7) & 0.030 & 0.030 & 0.952 (0.005)\\
\hspace{1em}npCBPS & -0.014 (0.001) & -15.2 (1.2) & 0.051 & 0.041 & 0.892 (0.007) & 0.003 (0.001) & 3.5 (0.7) & 0.030 & 0.029 & 0.950 (0.005)\\
\hspace{1em}GBM & -0.001 (0.001) & -1.0 (0.8) & 0.033 & 0.032 & 0.937 (0.005) & 0.003 (0.001) & 3.3 (0.7) & 0.030 & 0.030 & 0.947 (0.005)\\
\hspace{1em}Energy & -0.001 (0.001) & -1.3 (0.8) & 0.036 & 0.035 & 0.946 (0.005) & 0.002 (0.001) & 1.7 (0.8) & 0.034 & 0.034 & 0.948 (0.005)\\
\addlinespace[0.3em]
\multicolumn{11}{l}{\textbf{Poisson exposure}}\\
\hspace{1em}Unadjusted & 0.060 (0.001) & 63.3 (0.9) & 0.038 & 0.039 & 0.643 (0.011) & — & — & — & — & —\\
\hspace{1em}Adjusted & 0.000 (0.001) & -0.5 (1.0) & 0.042 & 0.043 & 0.944 (0.005) & — & — & — & — & —\\
\hspace{1em}Multinomial & -0.001 (0.001) & -1.5 (1.3) & 0.057 & 0.055 & 0.939 (0.005) & 0.008 (0.001) & 8.4 (1.1) & 0.048 & 0.048 & 0.945 (0.005)\\
\hspace{1em}CBPS & -0.001 (0.001) & -1.5 (1.3) & 0.057 & 0.054 & 0.944 (0.005) & 0.007 (0.001) & 7.2 (1.1) & 0.048 & 0.048 & 0.945 (0.005)\\
\hspace{1em}npCBPS & -0.025 (0.002) & -26.1 (2.4) & 0.102 & 0.073 & 0.901 (0.007) & 0.019 (0.001) & 19.7 (1.1) & 0.046 & 0.047 & 0.924 (0.006)\\
\hspace{1em}GBM & 0.000 (0.001) & -0.1 (1.4) & 0.059 & 0.056 & 0.940 (0.005) & 0.008 (0.001) & 8.3 (1.1) & 0.048 & 0.049 & 0.944 (0.005)\\
\hspace{1em}Energy & -0.001 (0.001) & -1.4 (1.4) & 0.060 & 0.060 & 0.945 (0.005) & 0.006 (0.001) & 6.3 (1.3) & 0.055 & 0.055 & 0.946 (0.005)\\
\bottomrule
\end{tabular}
\begin{tablenotes}
\item \textit{Note:} 
Entries are Monte Carlo summaries over 2,000 simulated datasets. Point estimates and confidence intervals were derived from weighted Poisson regression models with a sandwich estimator for variance. Unweighted outcome regressions (adjusted and unadjusted) are shown for comparison. All IPTW methods used stabilised weights. (np)CBPS \textendash{} (Non-Parametric) Covariate Balancing Propensity Scores; Emp. SE \textendash{} Empirical Standard Error; GBM \textendash{} Generalised Boosted Models; MCSE \textendash{} Monte Carlo Standard Error; Model SE \textendash{} Model-Based Standard Error; Rel. Bias \textendash{} Relative Bias.
\end{tablenotes}
\end{threeparttable}
\end{table}
\end{landscape}
}

\subsection{Properties of the weights}

\Cref{tab:sim_cc_balance} summarises covariate balance metrics and effective sample sizes for the unwinsorised weights. Multinomial, CBPS, and GBM weights resulted only in a small loss of precision: under the negative binomial DGM, mean ESS ranged from \num[round-precision = 0]{\simCcBalNegBinCbpsGbmMultiMinEss} to \num[round-precision = 0]{\simCcBalNegBinCbpsGbmMultiMaxEss}, with small standard deviations and 5\textsuperscript{th} percentile ESS all above \num[round-precision = 0]{\simCcBalNegBinCbpsGbmMultiFivePercEssMin}, close to the nominal 5{,}000 observations. Compared to these methods, energy balancing resulted in greater precision loss (mean ESS = \num[round-precision = 0]{\simCcBalNegBinEnergyRawMeanEss}). Similar patterns were observed under the Poisson DGM, although mean ESS was lower across all methods. npCBPS showed substantial and highly variable precision loss, with mean ESS of \num[round-precision = 0]{\simCcBalNegBinNpCbpsRawMeanEss} for the negative binomial and \num[round-precision = 0]{\simCcBalPoissonNpCbpsRawMeanEss} for the Poisson DGM, and very low ESS in some repetitions (for example, under the Poisson DGM the 5\textsuperscript{th} percentile ESS was \simCcBalPoissonNpCbpsRawFivePercEss).

Multinomial, CBPS, GBM, and energy weights achieved satisfactory exposure--covariate independence, with the mean absolute weighted correlation $|\rho_{w}| < 0.02$ for each method. Under the Poisson DGM, multinomial weights yielded larger residual correlations in some repetitions, indicated by $\max |\rho_{w}| = \num{\simCcBalPoissonMultinomialRawMaxW}$, although such cases were rare, with 95\textsuperscript{th} percentile $|\rho_{w}| = \num{\simCcBalPoissonMultinomialRawNineFiveRhoW}$. npCBPS achieved poorer balance compared to the other methods, with larger mean $|\rho_{w}|$ and more extreme tail values. Across methods, the average composite dependence metric $D_{w}$ and energy distance metrics $\epsilon_{A}$ and $\epsilon_{C}$ were close to zero, suggesting that the weighting generally preserved the marginal exposure and covariate distributions. However, these metrics were consistently largest for npCBPS, reinforcing its comparatively poorer balance and precision.

Winsorising the weights maintained greater precision, as reflected by a higher mean ESS, but this came at the cost of reduced covariate balance, evidenced by an increase in the average absolute exposure-covariate correlations.

\afterpage{
\begin{landscape}\begin{table}[htbp]
\scriptsize
\caption{\label{tab:sim_cc_balance}Covariate balance metrics for the simulations under complete data and RR = 1.1}
\begin{threeparttable}
\begin{tabular}[t]{lcccccccccccc}
\toprule
\textbf{Method} & \textbf{Mean ESS} & \textbf{SD ESS} & \textbf{5th perc ESS} & \textbf{95th perc ESS} & \textbf{Mean} $\boldsymbol{D_w}$ & \textbf{SD} $\boldsymbol{D_w}$ & \textbf{Mean} $\boldsymbol{\epsilon_A}$ & \textbf{Mean} $\boldsymbol{\epsilon_C}$ & \textbf{Mean} $\boldsymbol{|\rho_w|}$ & \textbf{SD} $\boldsymbol{|\rho_w|}$ & \textbf{95th} $\boldsymbol{|\rho_w|}$ & \textbf{Max} $\boldsymbol{|\rho_w|}$\\
\midrule
\addlinespace[0.3em]
\multicolumn{13}{l}{\textbf{Negative binomial exposure}}\\
\hspace{1em}Multinomial & 4,667 & 62 & 4,558 & 4,746 & <0.001 & <0.001 & <0.0001 & <0.0001 & 0.008 & 0.004 & 0.015 & 0.038\\
\hspace{1em}CBPS & 4,679 & 44 & 4,605 & 4,745 & <0.001 & <0.001 & <0.0001 & <0.0001 & <0.0001 & <0.0001 & <0.0001 & <0.0001\\
\hspace{1em}npCBPS & 3,243 & 838 & 1,717 & 4,412 & 0.006 & 0.004 & <0.001 & <0.001 & 0.044 & 0.029 & 0.093 & 0.138\\
\hspace{1em}GBM & 4,623 & 68 & 4,511 & 4,713 & 0.001 & <0.001 & <0.0001 & <0.0001 & 0.011 & 0.005 & 0.020 & 0.034\\
\hspace{1em}Energy & 3,639 & 64 & 3,527 & 3,737 & <0.001 & <0.0001 & <0.0001 & <0.0001 & 0.002 & <0.001 & 0.002 & 0.004\\
\addlinespace[0.3em]
\multicolumn{13}{l}{\textbf{Poisson exposure}}\\
\hspace{1em}Multinomial & 3,909 & 338 & 3,318 & 4,197 & <0.001 & 0.001 & <0.0001 & <0.0001 & 0.012 & 0.009 & 0.026 & 0.122\\
\hspace{1em}CBPS & 3,975 & 137 & 3,730 & 4,149 & <0.001 & <0.001 & <0.0001 & <0.0001 & <0.0001 & <0.0001 & <0.0001 & <0.0001\\
\hspace{1em}npCBPS & 1,675 & 1,007 & 453 & 3,589 & 0.012 & 0.010 & 0.001 & <0.001 & 0.064 & 0.044 & 0.147 & 0.233\\
\hspace{1em}GBM & 3,891 & 253 & 3,469 & 4,179 & 0.002 & <0.001 & <0.0001 & <0.001 & 0.016 & 0.007 & 0.029 & 0.059\\
\hspace{1em}Energy & 3,066 & 99 & 2,901 & 3,221 & <0.001 & <0.0001 & <0.0001 & <0.0001 & 0.003 & <0.001 & 0.004 & 0.005\\
\bottomrule
\end{tabular}
\begin{tablenotes}
\item \textit{Note:} 
The $\epsilon_C$ metric and the absolute treatment\textendash{}covariate correlations were averaged across the three confounders. (np)CBPS \textendash{} (Non-Parametric) Covariate Balancing Propensity Scores; $D_w$ \textendash{} distance metric optimised by the energy-balancing approach; $\epsilon_A$ \textendash{} energy distance between the weighted and unweighted marginal distributions of the exposure; $\epsilon_C$ \textendash{} energy distance between the weighted and unweighted marginal distributions of the covariates; ESS \textendash{} Effective Sample Size; GBM \textendash{} Generalised Boosted Models; $\rho_w$ \textendash{} average weighted treatment\textendash{}covariate correlation; perc \textendash{} percentile; SD \textendash{} standard deviation.
\end{tablenotes}
\end{threeparttable}
\end{table}
\end{landscape}
}

\subsection{Computational speed}
The methods varied in computational speed (Figure~\ref{fig:sim_relative_time} in the Supplementary Information), with the CBPS method being the fastest, with a mean absolute runtime of \compSpeedTimeCBPS\ seconds per run under the given local setup. Across the 20 simulated datasets, the multinomial binning approach was, on average, \compSpeedRelMultinomial\ times slower than CBPS. In contrast, npCBPS was approximately \compSpeedRelnpCBPS\ times slower, while GBM exhibited a relative slowdown of \compSpeedRelGBM\ times. Energy weights were particularly computationally intensive, requiring \compSpeedRelEnergy\ times longer than the multinomial weights.

\subsection{Performance under missing data}

Tables~\ref{tab:sim_miss_bias} and~\ref{tab:sim_miss_coverage} present performance of the `within-imputation' estimator. Under MCAR, the adjusted outcome model and the IPTW estimators based on multinomial, CBPS, GBM and energy weights maintained low bias even with a high proportion of missing data. When 60\% of rows were incomplete, relative bias for these methods remained within $\pm$3\% for the negative binomial exposure DGM and within about $\pm$5\% for the Poisson exposure DGM. Unadjusted estimation continued to display large positive bias (around 28\% for the negative binomial and 65\% for the Poisson exposure at 60\% missing). npCBPS showed persistent negative bias. Coverage under MCAR showed evidence of the pooled sandwich estimator being slightly conservative with higher missingness across methods.

\afterpage{
\begin{landscape}\begin{table}[htbp]

\caption{\label{tab:sim_miss_bias}Relative bias (MCSE) under varying missingness scenarios and RR = 1.1.}
\begin{threeparttable}
\begin{tabular}[t]{lccccccc}
\toprule
\multicolumn{2}{c}{\textbf{ }} & \multicolumn{3}{c}{\textbf{MCAR}} & \multicolumn{3}{c}{\textbf{MAR}} \\
\cmidrule(l{3pt}r{3pt}){3-5} \cmidrule(l{3pt}r{3pt}){6-8}
\textbf{Method} & \textbf{Complete} & \textbf{20\%} & \textbf{40\%} & \textbf{60\%} & \textbf{20\%} & \textbf{40\%} & \textbf{60\%}\\
\midrule
\addlinespace[0.3em]
\multicolumn{8}{l}{\textbf{Negative binomial exposure}}\\
\hspace{1em}Unadjusted & 27.1 (0.6) & 27.5 (0.7) & 28.0 (0.8) & 27.8 (0.9) & 22.1 (1.0) & 14.2 (1.3) & 7.7 (1.6)\\
\hspace{1em}Adjusted & -0.9 (0.6) & -0.4 (0.7) & 0.4 (0.8) & 0.5 (0.9) & -4.4 (1.0) & -10.4 (1.3) & -15.4 (1.7)\\
\hspace{1em}Multinomial & -1.0 (0.7) & -0.1 (0.8) & 1.2 (0.9) & 2.0 (1.0) & -3.0 (1.0) & -8.7 (1.4) & -14.1 (1.7)\\
\hspace{1em}CBPS & -2.2 (0.7) & -1.3 (0.8) & 0.0 (0.9) & 0.7 (1.0) & -3.8 (1.0) & -9.4 (1.3) & -14.4 (1.7)\\
\hspace{1em}npCBPS & -15.2 (1.2) & -14.4 (1.1) & -12.8 (1.1) & -11.1 (1.2) & -16.6 (1.3) & -21.8 (1.4) & -26.2 (1.7)\\
\hspace{1em}GBM & -1.0 (0.8) & -0.4 (0.8) & 0.7 (0.9) & 1.5 (1.0) & -2.8 (1.0) & -8.6 (1.3) & -13.8 (1.7)\\
\hspace{1em}Energy & -1.3 (0.8) & -0.4 (0.9) & 1.3 (0.9) & 2.7 (1.0) & -4.0 (1.1) & -9.8 (1.4) & -14.8 (1.7)\\
\addlinespace[0.3em]
\multicolumn{8}{l}{\textbf{Poisson exposure}}\\
\hspace{1em}Unadjusted & 63.3 (0.9) & 63.8 (1.0) & 64.7 (1.1) & 65.2 (1.3) & 60.1 (1.2) & 56.3 (1.5) & 50.8 (2.0)\\
\hspace{1em}Adjusted & -0.5 (1.0) & 0.5 (1.1) & 1.8 (1.2) & 3.3 (1.4) & -1.6 (1.3) & -2.7 (1.6) & -4.4 (2.1)\\
\hspace{1em}Multinomial & -1.5 (1.3) & -0.1 (1.4) & 1.7 (1.5) & 3.2 (1.6) & -1.6 (1.5) & -2.7 (1.8) & -4.6 (2.2)\\
\hspace{1em}CBPS & -1.5 (1.3) & -0.1 (1.4) & 1.2 (1.4) & 2.6 (1.6) & -1.5 (1.5) & -2.5 (1.8) & -4.6 (2.2)\\
\hspace{1em}npCBPS & -26.1 (2.4) & -22.1 (2.2) & -19.3 (2.1) & -19.0 (2.0) & -26.1 (2.2) & -26.4 (2.2) & -29.6 (2.3)\\
\hspace{1em}GBM & -0.1 (1.4) & 0.8 (1.4) & 2.0 (1.4) & 3.2 (1.6) & -0.0 (1.5) & -0.8 (1.7) & -2.7 (2.2)\\
\hspace{1em}Energy & -1.4 (1.4) & 0.6 (1.4) & 2.9 (1.5) & 4.9 (1.6) & -1.9 (1.5) & -2.8 (1.8) & -4.1 (2.2)\\
\bottomrule
\end{tabular}
\begin{tablenotes}
\item \textit{Note:} 
Entries are Monte Carlo summaries over 2,000 simulated datasets. Percentages under MCAR and MAR scenarios indicate the percentage of incomplete rows introduced. Numbers in brackets indicate Monte Carlo standard errors. Point estimates and confidence intervals were derived from weighted Poisson regression models with a sandwich estimator for variance. Unweighted outcome regressions (adjusted and unadjusted) are shown for comparison. For each repetition, results were pooled across multiply imputed datasets, with the number of imputations set to the percentage of incomplete cases. All weighting methods used stabilised weights, shown without winsorisation in this table. (np)CBPS \textendash{} (Non-Parametric) Covariate Balancing Propensity Scores; GBM \textendash{} Generalised Boosted Models; MAR \textendash{} Missing at Random; MCAR \textendash{} Missing Completely at Random; MCSE \textendash{} Monte Carlo Standard Error.
\end{tablenotes}
\end{threeparttable}
\end{table}
\end{landscape}
}

\afterpage{
\begin{landscape}\begin{table}[htbp]

\caption{\label{tab:sim_miss_coverage}Coverage (MCSE) under varying missingness scenarios and RR = 1.1.}
\begin{threeparttable}
\begin{tabular}[t]{lccccccc}
\toprule
\multicolumn{2}{c}{\textbf{ }} & \multicolumn{3}{c}{\textbf{MCAR}} & \multicolumn{3}{c}{\textbf{MAR}} \\
\cmidrule(l{3pt}r{3pt}){3-5} \cmidrule(l{3pt}r{3pt}){6-8}
\textbf{Method} & \textbf{Complete} & \textbf{20\%} & \textbf{40\%} & \textbf{60\%} & \textbf{20\%} & \textbf{40\%} & \textbf{60\%}\\
\midrule
\addlinespace[0.3em]
\multicolumn{8}{l}{\textbf{Negative binomial exposure}}\\
\hspace{1em}Unadjusted & 0.811 (0.009) & 0.824 (0.009) & 0.840 (0.008) & 0.860 (0.008) & 0.895 (0.007) & 0.927 (0.006) & 0.946 (0.005)\\
\hspace{1em}Adjusted & 0.953 (0.005) & 0.946 (0.005) & 0.941 (0.005) & 0.940 (0.005) & 0.943 (0.005) & 0.944 (0.005) & 0.957 (0.005)\\
\hspace{1em}Multinomial & 0.950 (0.005) & 0.950 (0.005) & 0.955 (0.005) & 0.952 (0.005) & 0.950 (0.005) & 0.949 (0.005) & 0.962 (0.004)\\
\hspace{1em}CBPS & 0.951 (0.005) & 0.954 (0.005) & 0.952 (0.005) & 0.953 (0.005) & 0.952 (0.005) & 0.950 (0.005) & 0.961 (0.004)\\
\hspace{1em}npCBPS & 0.892 (0.007) & 0.940 (0.005) & 0.962 (0.004) & 0.970 (0.004) & 0.950 (0.005) & 0.969 (0.004) & 0.973 (0.004)\\
\hspace{1em}GBM & 0.937 (0.005) & 0.953 (0.005) & 0.958 (0.005) & 0.959 (0.004) & 0.960 (0.004) & 0.958 (0.004) & 0.964 (0.004)\\
\hspace{1em}Energy & 0.946 (0.005) & 0.959 (0.004) & 0.971 (0.004) & 0.969 (0.004) & 0.962 (0.004) & 0.965 (0.004) & 0.972 (0.004)\\
\addlinespace[0.3em]
\multicolumn{8}{l}{\textbf{Poisson exposure}}\\
\hspace{1em}Unadjusted & 0.643 (0.011) & 0.678 (0.010) & 0.709 (0.010) & 0.765 (0.009) & 0.790 (0.009) & 0.839 (0.008) & 0.897 (0.007)\\
\hspace{1em}Adjusted & 0.944 (0.005) & 0.950 (0.005) & 0.946 (0.005) & 0.944 (0.005) & 0.951 (0.005) & 0.948 (0.005) & 0.940 (0.005)\\
\hspace{1em}Multinomial & 0.939 (0.005) & 0.960 (0.004) & 0.966 (0.004) & 0.967 (0.004) & 0.956 (0.005) & 0.966 (0.004) & 0.959 (0.004)\\
\hspace{1em}CBPS & 0.944 (0.005) & 0.959 (0.004) & 0.964 (0.004) & 0.966 (0.004) & 0.957 (0.005) & 0.962 (0.004) & 0.961 (0.004)\\
\hspace{1em}npCBPS & 0.901 (0.007) & 0.959 (0.004) & 0.969 (0.004) & 0.979 (0.003) & 0.954 (0.005) & 0.967 (0.004) & 0.976 (0.003)\\
\hspace{1em}GBM & 0.940 (0.005) & 0.962 (0.004) & 0.968 (0.004) & 0.972 (0.004) & 0.968 (0.004) & 0.970 (0.004) & 0.967 (0.004)\\
\hspace{1em}Energy & 0.945 (0.005) & 0.974 (0.004) & 0.982 (0.003) & 0.980 (0.003) & 0.965 (0.004) & 0.971 (0.004) & 0.970 (0.004)\\
\bottomrule
\end{tabular}
\begin{tablenotes}
\item \textit{Note:} 
Entries are Monte Carlo summaries over 2,000 simulated datasets. Percentages under MCAR and MAR scenarios indicate the percentage of incomplete rows introduced. Numbers in brackets indicate Monte Carlo standard errors. Point estimates and confidence intervals were derived from weighted Poisson regression models with a sandwich estimator for variance. Unweighted outcome regressions (adjusted and unadjusted) are shown for comparison. For each repetition, results were pooled across multiply imputed datasets, with the number of imputations set to the percentage of incomplete cases. All weighting methods used stabilised weights, shown without winsorisation in this table. (np)CBPS \textendash{} (Non-Parametric) Covariate Balancing Propensity Scores; GBM \textendash{} Generalised Boosted Models; MAR \textendash{} Missing at Random; MCAR \textendash{} Missing Completely at Random; MCSE \textendash{} Monte Carlo Standard Error.
\end{tablenotes}
\end{threeparttable}
\end{table}
\end{landscape}
}

Under MAR, the `within-imputation' estimator showed increasing negative bias as the amount of missingness increased, particularly under the negative binomial DGM. At 20\% of incomplete rows, multinomial, CBPS, and energy weights showed a small relative bias between -3.0\% and -4.0\%, depending on method, which then increased to between -13.8\% and -14.8\% at 60\% of incomplete rows. However, this increase in bias was also apparent for regression adjustment, demonstrating that it was not specifically due to the performance of the weighting estimators. While a similar pattern was observed under the Poisson DGM, relative bias was smaller, ranging from -4.1\% to -4.6\% across the three weighting methods. For the unadjusted method, negative bias from underadjustment partly counterbalanced the positive bias from increasing missingness, resulting in lower net relative bias at higher missingness. This bias cancellation should not be interpreted as evidence of superior performance under higher missingness. npCBPS remained the method with highest bias among the weighting estimators across all DGMs with missingness. Coverage under MAR showed similar patterns to those under MCAR.

The increased bias under MAR was not specific to the weighting methods, as it was observed for both regression adjustment and IPTW. We therefore hypothesised that this bias may have arisen from the imputation procedure, particularly from imputing the count exposure $A$, since it was more pronounced under the negative binomial exposure than under the Poisson DGM. To examine this, we conducted an additional sensitivity analysis in which the same overall proportion of missingness (20\%, 40\%, and 60\%) was imposed while keeping $A$ fully observed. Thus, $C_1$ and $A$ were fully observed, whereas MAR missingness was introduced only in $C_2$, $C_3$, and $Y$, using the same calibration procedure. Owing to the substantial computational burden, this analysis was restricted to the negative binomial data-generating mechanism with true $\mathrm{RR} = 1.1$ and to regression adjustment, CBPS, and multinomial weights. Under this setting, all three methods showed low relative bias and acceptable coverage (Table~\ref{tab:sim_noexpmiss}).

\section{Application to the motivating example}\label{sec4}

We now apply the studied methods to the motivating example to estimate the effect of psychological distress at age 34 on the presence of a longstanding illness at age 42. As in the simulations, the RR was estimated using modified Poisson regression re-weighted using the outlined IPTW approaches, except for energy balancing, which was omitted from this analysis since it imposed excessive computational demands. We also provide results for the unweighted, unadjusted model and the covariate-adjusted model for comparison. To aid interpretability, the RR was rescaled per 4-point higher Malaise Inventory score, as a change from 0 to 4 points is regarded as reflecting a shift from minimal symptoms to reaching the threshold for psychiatric caseness.\cite{arias-de_la_torre_depressive_2021, gondek_psychological_2022}

Missing data were handled using MI by chained equations, which was combined with IPTW using the previously outlined `within-imputation' approach. The imputation models included all variables from the substantive models, as well as selected auxiliary variables likely related to attrition and the underlying missing values. The selection of auxiliary variable was based on prior work \cite{katsoulis_data_2024} and is further described in the Supplementary Information. As in the simulations, the number of imputed datasets was set equal to the percentage of incomplete rows in the dataset ($m$ = \num[round-precision = 0]{\bcsM} for this motivating example).

The dataset included \num[round-precision = 0]{\bcsEligible} eligible participants (born in England, Scotland, or Wales, remained alive and did not emigrate by age 42). Descriptive statistics of the sample are provided in the Supplementary Information (Table~\ref{tab:descriptives}). Just under half of the sample were female, with 66.6\% married or partnered. At age 30, 23.0\% reported having a longstanding illness, and this increased to 29.1\% by age 42.

\begin{figure}[t]
  \centering
  \includegraphics[width=\linewidth]{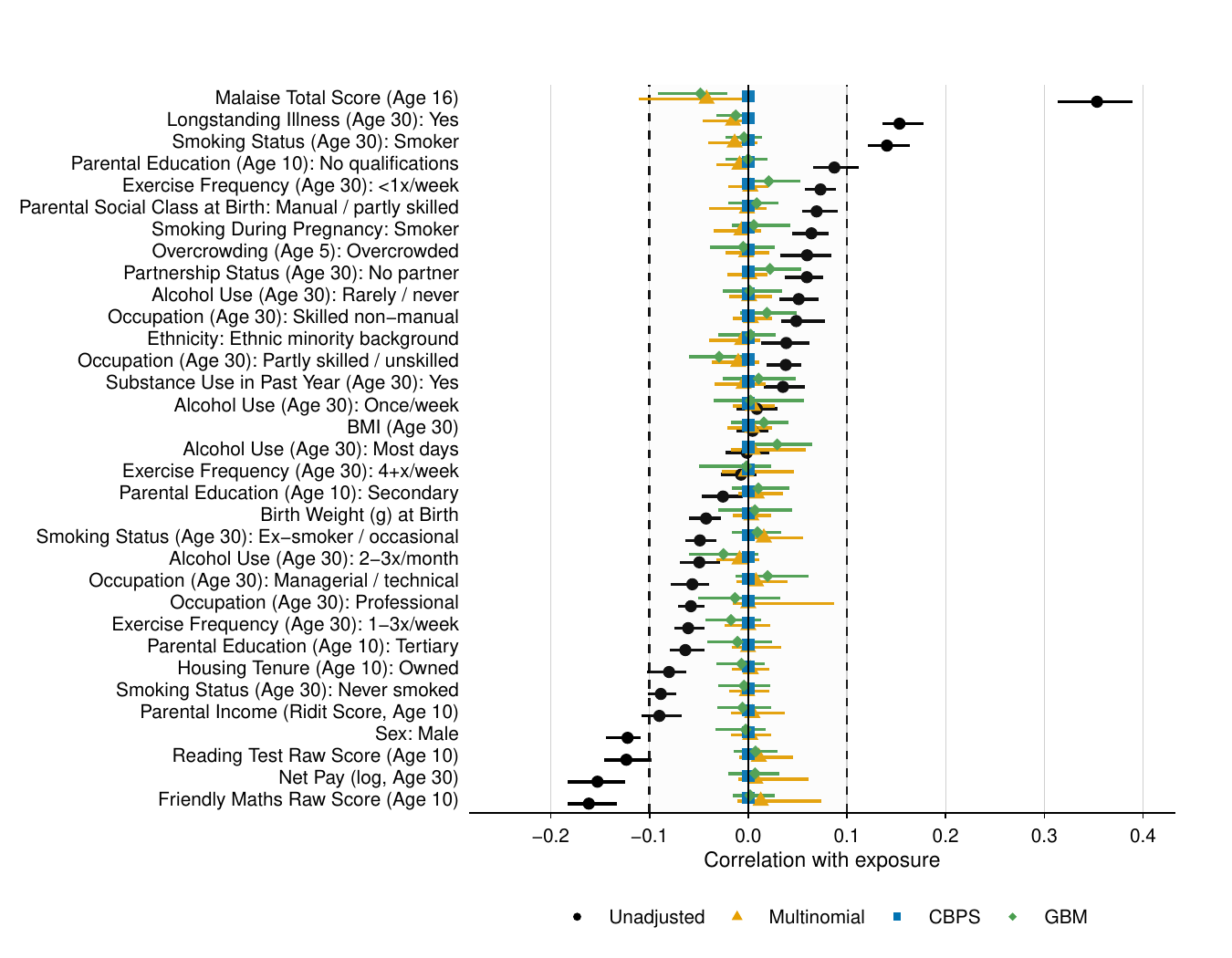}
  \caption{Balance plot for the motivating example. Points show weighted or unadjusted Pearson and point-biserial correlations between the exposure and each covariate; horizontal whiskers show the range of the correlations across \num[round-precision = 0]{\bcsM} multiply imputed datasets. The npCBPS method failed to achieve balance for several covariates and produced wide correlation ranges, so its results are omitted here for visual clarity; a corresponding balance plot including all methods is provided in Figure~\ref{fig:bcs70_balance_npcbps} of the Supplementary Information. Data are from the 1970 British Cohort Study (N = \num[round-precision = 0]{\bcsEligible}). (np)CBPS -- (Non-Parametric) Covariate Balancing Propensity Score; GBM -- Generalised Boosted Models.}
  \label{fig:bcs70_balance}
\end{figure}

Figure~\ref{fig:bcs70_balance} summarises covariate balance before and after weighting in terms of weighted correlations between the exposure and each covariate across the \num[round-precision = 0]{\bcsM} multiply imputed datasets. In the unweighted data, the
Malaise score showed an association of $|\rho|$ > 0.1 with a number of covariates, including the past levels of psychological distress, childhood cognition, own income, parental income, sex at birth, past longstanding illness, and smoking status. Weighting using multinomial models, CBPS, and GBM succeeded in reducing most correlations to values close to zero and produced relatively tight ranges across imputations, indicating good balance. By contrast, npCBPS failed to balance several covariates and yielded much wider ranges of correlations across imputations, consistent with unstable weights (Figure~\ref{fig:bcs70_balance_npcbps} in the Supplementary Information). Based on this assessment, we would refrain from using npCBPS-derived weights further for effect estimation, but here we provide this effect estimate regardless for illustrative purposes.

\begin{table}[t]
\centering
\begin{threeparttable}
\caption{\label{tab:bcs70_balance}Covariate balance metrics for the BCS70 motivating example}
\begin{tabular}[t]{lccccc}
\toprule
\textbf{Method} & \textbf{Mean \boldmath$|\mathrm{\rho_w}|$} & \textbf{\boldmath$\%\,|\mathrm{\rho_w}|>0.10$} & \textbf{Max \boldmath$|\mathrm{\rho_w}|$\textsuperscript{\dag}} & \textbf{Mean ESS} & \textbf{5\textsuperscript{th} pct ESS}\\
\midrule
\addlinespace[0.3em]
\multicolumn{6}{l}{\textbf{Raw weights}}\\
\hspace{1em}Multinomial & 0.010 & 0.1 & 0.111 & 9,690 & 6,294\\
\hspace{1em}CBPS & \textless{}0.001 & 0 & \textless{}0.001 & 11,868 & 11,199\\
\hspace{1em}GBM & 0.015 & 0 & 0.092 & 10,087 & 8,517\\
\hspace{1em}npCBPS & 0.027 & 4.3 & 1.286 & 6,956 & 457\\
\addlinespace[0.3em]
\multicolumn{6}{l}{\textbf{Winsorised weights}}\\
\hspace{1em}Multinomial & 0.009 & 0 & 0.047 & 13,698 & 13,418\\
\hspace{1em}CBPS & 0.013 & 0 & 0.076 & 14,248 & 13,950\\
\hspace{1em}GBM & 0.018 & 0 & 0.063 & 13,574 & 13,108\\
\hspace{1em}npCBPS & 0.032 & 3.0 & 0.170 & 15,688 & 15,333\\
\bottomrule
\end{tabular}
\vspace{0.3em}
\par\small
\textit{Note:} Balance metrics are summarised across all covariates and imputed data sets. Weighted correlation is computed as the weighted covariance between exposure and covariate divided by the product of unweighted standard deviations of the exposure and covariate. ESS -- effective sample size; $\mathrm{\rho_w}$ -- weighted correlation between covariate and exposure.

\smallskip
\hangindent=1.2em\hangafter=1
\textsuperscript{\dag}\,Values can exceed 1 when weights are extreme/unstable. This occurs because the weighted covariance is scaled by unweighted standard deviations; this is to ensure that lower values indicate superior balance.
\end{threeparttable}
\end{table}

\begin{figure}[t]
  \centering
  \includegraphics[width=\linewidth]{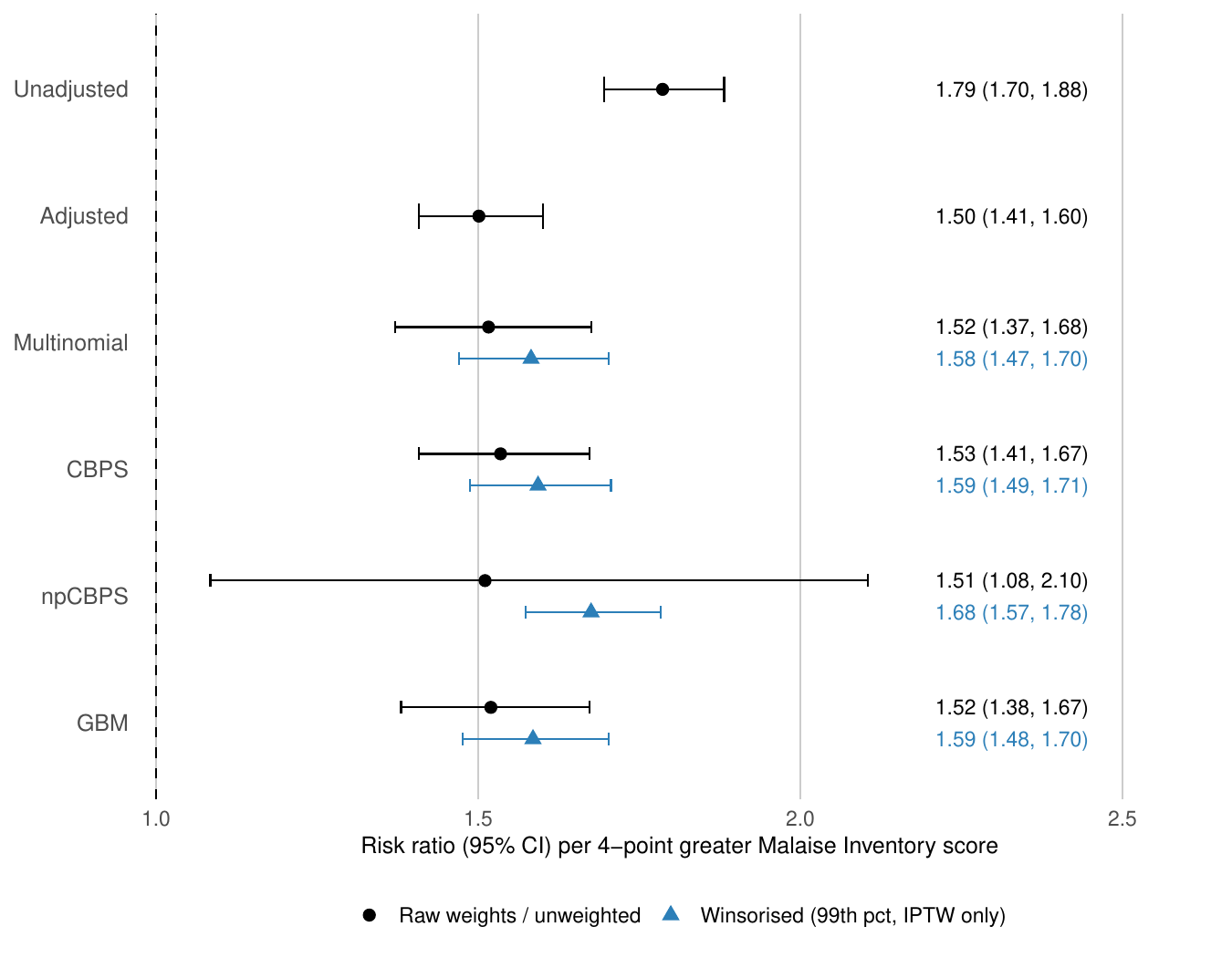}
  \caption{The forest plot shows risk ratios and 95\% confidence intervals for longstanding illness per 4-point higher Malaise Inventory score from unadjusted, inverse probability of treatment weighted, and covariate-adjusted Poisson regression models, pooled across \num[round-precision = 0]{\bcsM} multiply imputed datasets (sample N = \num[round-precision = 0]{\bcsEligible}). The unadjusted model gives the largest estimate (risk ratio 1.79); all weighted and covariate-adjusted analyses give similar risk ratios (ranging from 1.50--1.53 across the methods). Winsorisation of weights at the 99\textsuperscript{th} percentile yielded slightly higher estimates than the corresponding raw weights. CI widths are similar across methods, except for npCBPS, where the wide CI (1.08--2.10) reflects extreme weights and is substantially reduced after winsorisation. (np)CBPS -- (Non-Parametric) Covariate Balancing Propensity Scores; GBM -- Generalised Boosted Models.}
  \label{fig:bcs70_forest_combined}
\end{figure}

Estimated associations between psychological distress and longstanding illness are shown in Figure~\ref{fig:bcs70_forest_combined}. In the unadjusted Poisson regression model, a 4-point higher Malaise score at age 34 was associated with a risk ratio of \bcsResrawunadjustedestimate{} (95 \% confidence interval \bcsResrawunadjustedconfLow, \bcsResrawunadjustedconfHigh) for longstanding illness at age 42. All covariate-adjusted and inverse probability weighted analyses yielded attenuated but broadly similar risk ratios of around \bcsResrawadjustedestimate--\bcsResrawCBPSestimate, with substantial overlap of confidence intervals across methods. The npCBPS estimator produced a comparable point estimate but with a much wider confidence interval (\bcsResrawnpCBPSconfLow, \bcsResrawnpCBPSconfHigh), reflecting the extreme weights observed in the balance diagnostics; winsorising the weights at the 99th percentile narrowed this interval, with a slightly higher point estimate. Overall, in this applied example, the different weighting approaches (excluding npCBPS) and conventional covariate adjustment give comparable estimates for the effect of psychological distress on longstanding illness in mid-life.

\FloatBarrier
\section{Discussion}\label{sec5}

Using simulated data informed by a motivating example, we demonstrated that several IPTW approaches can be applied to estimate causal effects of count exposures on binary outcomes. Under conditions of complete data and linear covariate effects, CBPS, GBM, multinomial weights, and energy weights produced estimates with no or minimal bias, and the sandwich variance estimator yielded near-nominal coverage. While all four methods appeared suitable under the simulated setting, selecting the approach may depend on the trade-offs between statistical performance, computational burden, and ease of implementation in a given application. Among the tested methods, CBPS consistently performed well across all metrics and preserved the highest level of precision among the weighting methods. Multinomial weights were also fast to compute and generally performed well; this method is also conceptually simpler compared to the other approaches, and therefore may be of interest to those with limited experience of complex IPTW algorithms. GBM also showed favourable performance in terms of bias, coverage, and balance, but required more computation time than CBPS and multinomial weights and may require further tuning in more complex settings.\cite{griffin_chasing_2017} Energy weights also produced estimates with low bias but at a substantially higher computational cost and with a greater loss of precision.

Weights estimated via npCBPS performed poorly on most metrics across all considered DGMs. This result was unexpected given the method's appeal, namely that it does not require specifying the marginal or conditional exposure distribution, and because a prior simulation study reported performance similar to parametric CBPS and binning-based approaches.\cite{sack_inverse_2023} In our simulations, npCBPS frequently produced extreme weights, as reflected in low ESS, and this is a plausible explanation for the observed bias and under-coverage. Winsorisation substantially improved npCBPS performance, with marked gains in ESS and coverage, but results remained inferior to those of the other methods. One possible explanation is that, under the skewed and truncated exposure distributions considered here, the parametric restrictions in CBPS constrained estimation away from highly extreme weighting solutions, whereas npCBPS could converge to more extreme weights. This interpretation is consistent with poorer npCBPS performance under the negative binomial mechanism, where truncation and skewness were more pronounced. Overall, we do not recommend using npCBPS for count exposures due to its tendency to generate extreme weights. More generally, these findings highlight that despite its theoretical appeal, researchers should not default to npCBPS when a treatment distribution is difficult to specify, and weight diagnostics (e.g. covariate balance assessment, ESS) and potential mitigation strategies (e.g. winsorisation or truncation) should be carefully considered.\cite{austin_assessing_2019}

Winsorisation of weights at the 99th percentile had only modest effects for the well-performing methods (CBPS, multinomial, GBM, and energy weights), with minimal changes in point estimates and coverage. These small shifts were consistent with the usual bias--variance trade-off \cite{cole_constructing_2008, lee_weight_2011}: modest gains in precision sometimes came at the cost of slightly poorer balance and small increases in bias. These results suggest that winsorisation is not necessary when the weight distribution is stable and ESS is not far from its nominal value, although it can be useful when extreme weights are present. Winsorisation may also serve as a sensitivity analysis to assess whether results are driven by a small number of extreme weights.

When combining IPTW with MI using the `within-imputation' approach, performance under MCAR was reassuring. Across missingness levels up to 60\% incomplete rows, IPTW using multinomial, CBPS, GBM, and energy weights maintained low bias, comparable in magnitude to that of the covariate-adjusted outcome model, while Rubin's rules applied to the sandwich variance estimator showed only slight conservatism at higher missingness with acceptable coverage overall. npCBPS continued to display persistent negative bias, consistent with its behaviour under complete data.

Under MAR, bias increased with higher missingness for both the weighted and covariate-adjusted estimators, particularly under the truncated negative binomial exposure mechanism. The fact that IPTW estimators (multinomial, CBPS, GBM, and energy weights) were no more biased than regression adjustment suggests that the additional bias under MAR was not driven primarily by the weighting procedures. This interpretation was reinforced by our sensitivity analysis, where the exposure was kept fully observed while MAR missingness was imposed in the remaining variables: under this setting, both regression adjustment and IPTW (multinomial and CBPS weights) performed well, with low bias and acceptable coverage. The MAR results for the Poisson exposure were less concerning: while bias increased somewhat with missingness, its magnitude remained relatively modest even at 60\% incomplete rows and was substantially smaller than under the truncated negative binomial mechanism. Taken together, these findings suggest that IPTW can still be used in this context when combined with MI, and that the main source of additional bias under MAR in our simulations was likely the imputation of the partially observed count exposure.

A plausible contributor to bias under MAR might be misspecification or limited suitability of the imputation model, specifically the use of predictive mean matching to impute positively skewed counts. Predictive mean matching restricts imputations to values within the range observed in the incomplete dataset. When missingness affects sparse regions of the covariate-pattern space (for example, the upper tail), suitable donors can be scarce or absent. In such cases, recovery of the missing information would require extrapolation beyond observed values among donors, which predictive mean matching does not provide. \cite{von_hippel_imputing_2025} In our negative binomial setting, right skewness leaves higher counts sparsely represented, which can limit the availability of suitable predictive mean matching donors and increase imputation error as missingness rises. This interpretation is consistent with fewer issues under the Poisson exposure, where truncation was less pronounced. Model-based alternatives for imputing count data have been developed, including approaches that accommodate zero-truncation and overdispersion. \cite{kleinke_multiple_2013} In principle, such approaches could mitigate donor-availability problems because they generate values from an estimated count distribution rather than recycling observed donor values. However, in our setting the imputation model would need to accommodate both overdispersion and right truncation, which, to our knowledge, remains an underexplored area. Given the lack of methods suitable for this setting, our results under MAR should be interpreted as reflective of current practical constraints of MI, rather than a limitation of IPTW for count exposures or the `within-imputation' approach to combining IPTW with MI.

When deciding on the best approach for a given situation, additional points and limitations of this study should be considered. As an initial evaluation of these methods in this setting, we focused on a restricted set of scenarios. The computational burden of some methods, particularly energy balancing, further limited the range of additional scenarios that could be explored within the scope of this study. First, the data-generating mechanisms were deliberately simple: covariate effects were linear, truncation of the count exposure was mild, and there was no extreme zero inflation. More complex distributions or stronger deviations from the motivating example might yield different patterns of performance. This consideration is especially important with respect to performance of energy balancing and GBMs, as performance comparisons under simple DGMs may fail to showcase their full benefits.\cite{setodji_right_2017, huling_independence_2024} Secondly, the covariate space was low-dimensional; multinomial binning in particular may not scale well to high-dimensional settings, where sparsity may become more pronounced. Thirdly, we considered only one type of MAR structure and one pattern of increasing missingness across variables. Fourthly, our simulations were conducted in relatively large samples. Performance, especially of methods prone to extreme weights, may degrade further in smaller samples.\cite{petersen_diagnosing_2012, zhou_propensity_2020} We also did not evaluate the broader range of available methods for count or numeric exposures, including other balance-targeting approaches such as covariate association eliminating weights\cite{yiu_covariate_2018} and entropy-balancing methods\cite{tubbicke_entropy_2022}, as well as other machine learning algorithms. \cite{chipman_bart_2010, kreif_evaluation_2015} These are important comparators for future work. Finally, we mostly relied on default tuning parameters as implemented in the \texttt{WeightIt} package, which are likely to reflect typical practice but may be sub-optimal in some scenarios. More customised tuning might improve performance for some methods at the cost of additional complexity.

\section{Conclusions}\label{sec:conclusions}

This simulation study suggests that several IPTW approaches can be used to estimate causal effects of count exposures in relatively large samples with moderate overdispersion and limited truncation, provided that exposure distributions and covariate structures do not depart too far from those considered here. Stabilised weights may be constructed using multinomial models with binning, CBPS, GBMs, or energy weights, with methods offering different benefits and trade-offs. In contrast, npCBPS performed poorly and is not recommended in these settings. With any weighting method, we recommend verifying that covariate balance has been achieved. IPTW can also be combined with MI in this setting, applying Rubin's rules to pool sandwich variance, although our results highlight the need for further research into imputation methods for right-truncated overdispersed counts.

\section*{Acknowledgments}

The authors acknowledge the use of the UCL Myriad High Performance Computing Facility (Myriad@UCL), and associated support services, in the completion of this work.

\section*{Funding}

MND is funded by the ESRC-BBSRC Social-Biological (Soc-B) Centre for Doctoral Training (ES/P000347/1). JKB is funded by a National Institute for Health and Care Research (NIHR) Advanced Fellowship (NIHR305289). The UCL Centre for Longitudinal Studies is supported by the Economic and Social Research Council [grant number ES/W013142/1]. The views expressed are those of the authors and not necessarily those of the ESRC, BBSRC, NIHR or the Department of Health and Social Care.

\section*{Conflict of interest}

The authors declare no potential conflict of interests.

\section*{Data Availability Statement}

The R code used for the analyses in this paper is available in dedicated GitHub repositories, with the repository for the simulation study located at \url{https://github.com/martindanka/iptw-sim-core}  and the repository demonstrating the application of the methods to the motivating example available at \url{https://github.com/martindanka/iptw-sim-example}. Data from the 1970 British Cohort Study can be obtained from the UK Data Service under a standard End User Licence Agreement (Study Number 200001).


\appendix
\renewcommand{\thefigure}{S\arabic{figure}}
\renewcommand{\thetable}{S\arabic{table}}
\setcounter{figure}{0}
\setcounter{table}{0}

\clearpage
\section*{Supplementary Information}
\addcontentsline{toc}{section}{Supplementary Information}

\subsection*{Generating missingness}

Missingness was introduced in four variables, in fixed order: $C_2$, $C_3$, $A$, and $Y$. The binary covariate $C_1$
was kept fully observed. The mechanism definitions are given in Section~\ref{sec:simdesign}. Here we report only the calibration and
implementation details used in code.

For both mechanisms, we parameterised per-variable missingness as
$p_i = p_0 + (i-1)d,\ i=1,\dots,4$, with constraints
$p_0 \ge 0$, $d \ge 0$, and $p_0 + 3d < 1$, and target values
$\phi^\ast \in \{0.20,0.40,0.60\}$. Calibration used \texttt{constrOptim()}
in R (barrier-constrained optimisation; \texttt{maxit = 1000}) with initial
values $p_0=d=\max(\phi^\ast/10,10^{-4})$.

\textbf{MCAR implementation.}
For MCAR, each objective evaluation was analytical:
$\phi_{\text{MCAR}} = 1-\prod_{i=1}^4(1-p_i)$.
The optimiser therefore minimised
$\{\phi_{\text{MCAR}}-\phi^\ast\}^2$ directly, then Bernoulli masks were drawn
independently for $C_2$, $C_3$, $A$, and $Y$ using the calibrated $p_i$.

\textbf{MAR implementation.}
For MAR, calibration was done separately by DGM (Poisson and negative binomial)
on a fixed complete reference sample of size $N=1{,}000{,}000$.
For each candidate $(p_0,d)$ in the outer optimisation:
\begin{enumerate}
\item Set $p_i = p_0 + (i-1)d$.
\item Compute linear predictors from the parent graph
$C_2 \leftarrow C_1$,
$C_3 \leftarrow (C_1,C_2)$,
$A \leftarrow (C_1,C_2,C_3)$,
$Y \leftarrow (C_1,C_2,C_3,A)$,
with unit coefficients for all parent terms.
\item For each variable, solve for $\gamma_{0,i}$ from
$\frac{1}{N}\sum_{j=1}^{N}\text{expit}(\gamma_{0,i}+\eta_{ij})-p_i=0$
using \texttt{uniroot()} on $[-20,20]$
(Brent-type root finding in base R).
\item Evaluate
$\hat\phi_{\text{MAR}}=\frac{1}{N}\sum_{j=1}^{N}\left[1-\prod_{i=1}^{4}
\{1-\text{expit}(\gamma_{0,i}+\eta_{ij})\}\right]$.
\item Return $\left(\hat\phi_{\text{MAR}}-\phi^\ast\right)^2$ to the outer
\texttt{constrOptim()} routine.
\end{enumerate}

The optimised $(p_0,d)$, derived $p_i$, and $\gamma_{0,i}$ values were stored
and then reused during simulation runs. In each simulated dataset, missingness
indicators were generated as
$M_{ij}\sim\text{Bernoulli}[\text{expit}(\gamma_{0,i}+\eta_{ij})]$, and values
with $M_{ij}=1$ were set to missing. Because the reference dataset was fixed
during calibration, the procedure was deterministic for each DGM and target
$\phi^\ast$. The Monte Carlo component arose only from approximating
expectations on the large reference sample. Final calibrated values used in the
simulation are reported in Table~\ref{tab:missingness_parameters}.

\subsection*{Simulation implementation on HPC}

The simulation runs were executed on UCL Myriad using Sun Grid Engine (SGE) with R v4.4.0. The complete-case scenarios were run for all three effect sizes (RR = 1.0, 1.1, and 1.2), while the missing-data scenarios (MCAR and MAR) were run for RR = 1.1 only, yielding 18 simulation scenarios in total, each with 2,000 replications (36,000 individual tasks).

Parallelisation was implemented through SGE job arrays, with one single-core array task per replication. Because computational burden varied across scenarios, particularly between complete-case and missing-data designs requiring multiple imputation, resource requests were tuned in memory/wall-clock tiers (5--8\,GB RAM and 00:20:00--06:30:00 wall-clock time). Each task copied the project to a node-local temporary directory before execution to reduce shared filesystem I/O contention. While absolute computation time depends on the cluster hardware, job scheduling, and queue availability, this full set of simulations took on the order of one week to complete on UCL Myriad, counted from submission.

For reproducibility, random-number generation used the L'Ecuyer Combined Multiple Recursive Generator (CMRG) with a fixed base seed. Task-specific substreams were assigned by advancing the generator to the substream indexed by the array task ID (\texttt{parallel::nextRNGSubStream()}), ensuring non-overlapping random streams across replications.\cite{lecuyer_combined_1996, morris_using_2019}

R package dependencies were managed via project-level \texttt{renv} lockfiles, with separate lockfiles for the HPC environment (R v4.4.0, Linux) and the local analysis environment (R v4.4.1, macOS). Local analyses, including computation speed benchmarks, result processing, and the motivating example, were conducted using the local environment. The full codebase needed to re-run the simulations as well as local processing is provided in the main project GitHub repository (\url{https://github.com/martindanka/iptw-sim-core}).

\subsection*{Computation speed}

To compare computational performance, we simulated 20  datasets as described in the manuscript, using the negative binomial exposure data-generating mechanism. For each dataset and each weighting method, we recorded the CPU time of weight estimation, denoted $T_{m, d}$. For the multinomial method, exposure categories with prevalence <1\% were first collapsed and this preprocessing step was included in the timed computation, as it is required specifically for that method.

\begin{figure}[H]
  \centering
  \includegraphics[width=\textwidth]{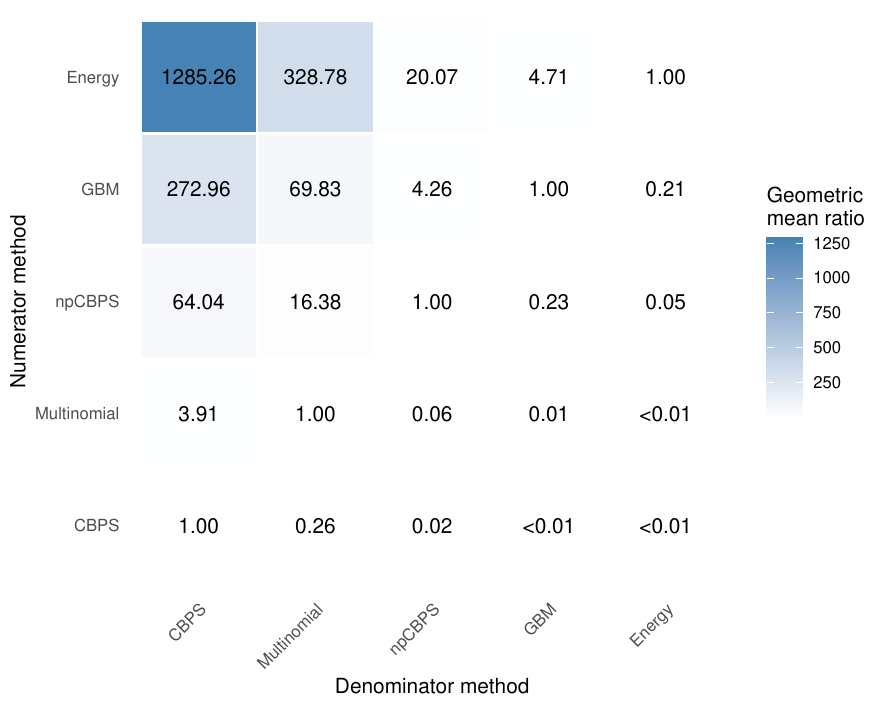}
  \caption{A heatmap comparing relative computation times of the five weight estimation methods. Pairwise relative times (ratios of CPU times) were obtained for each of the 20 simulated datasets, each containing N = 5,000 observations. Geometric means of these ratios were taken across all datasets for each pairwise comparison. A higher relative time indicates that the numerator method was slower against the method in the denominator. All methods were implemented using the WeightIt R package. The CPU time of the multinomial approach included preprocessing of the exposure by collapsing categories with prevalences below 1\%. (np)CBPS, (Non-Parametric) Covariate Balancing Propensity Score; GBM, Generalised Boosted Models.}
  \label{fig:sim_relative_time}
\end{figure}

Within each dataset, we then compared methods pairwise by calculating $T_{m,d}/{T_{m',d}}$, that is, the runtime of one method divided by the runtime of another on the same dataset. This produced a pairwise relative-time matrix $R^{d}$ for each dataset $d$, where each entry is the ratio $R_{m, m'}^{d} = T_{m,d}/{T_{m',d}}$. To summarise performance across simulations, we averaged the log runtime ratio for each pair of methods across the 20 datasets and exponentiated these results back to the original scale\cite{fleming_how_1986}:
\[
\bar{R}_{m,m'} = \exp\left(\frac{1}{20}\sum_{d=1}^{20} \log R^{(d)}_{m,m'}\right)
\]
Each cell of the resulting heatmap therefore represents the geometric mean runtime ratio for a given ordered pair of methods, computed across datasets. This metric therefore provides information about how much longer one method typically takes relative to another for a dataset generated under this setting. The resulting heatmap is available in Figure~\ref{fig:sim_relative_time}.

All methods were benchmarked on a local machine, using the same 20 simulated datasets generated under a fixed random seed. Before benchmarking, we ran each method once on the first dataset as a warm-up step so that one-off start-up costs, such as first-use loading and initial compilation overheads, occurred before timing began. CPU time was recorded using a common timing routine. Within each dataset, methods were executed in a randomised order, with garbage collection performed before each method call to reduce carry-over memory effects. Software details and hardware specifications were recorded in the same session and can be accessed in the GitHub repository (\url{https://github.com/martindanka/iptw-sim-core}).

\subsection*{Auxiliary variables in motivating example MI}

For the motivating example, MI included all substantive-model variables and additional
auxiliary variables informed by predictors of non-response in BCS70, identified in previous work.\cite{katsoulis_data_2024}
Specifically, this was based on (i) predictors of non-response at age 42 and (ii) predictors that were consistently associated with non-response across ages
26--46. Some of these variables or their closely related constructs were already present in the substantive model as potential confounders,
so these were not added as separate auxiliary-only variables.

Additional auxiliary variables included:
\begin{itemize}
\item cumulative indicators of prior non-response across previous sweeps (participation history);
\item number of antenatal visits (birth sweep);
\item certainty of the date of the last menstrual period (birth sweep);
\item copying design test score (age 5--6 sweep);
\item voting in the 1997 general election (age 30 sweep);
\item working overtime (age 34 sweep);
\item endorsement of the statement that everyone should behave responsibly (age 34 sweep);
\item having been found guilty in a criminal court (age 34 sweep);
\item willingness to be contacted for the parents' research project (age 38 sweep).
\end{itemize}


\clearpage

\begin{figure}[H]
  \centering
  \includegraphics[width=\textwidth]{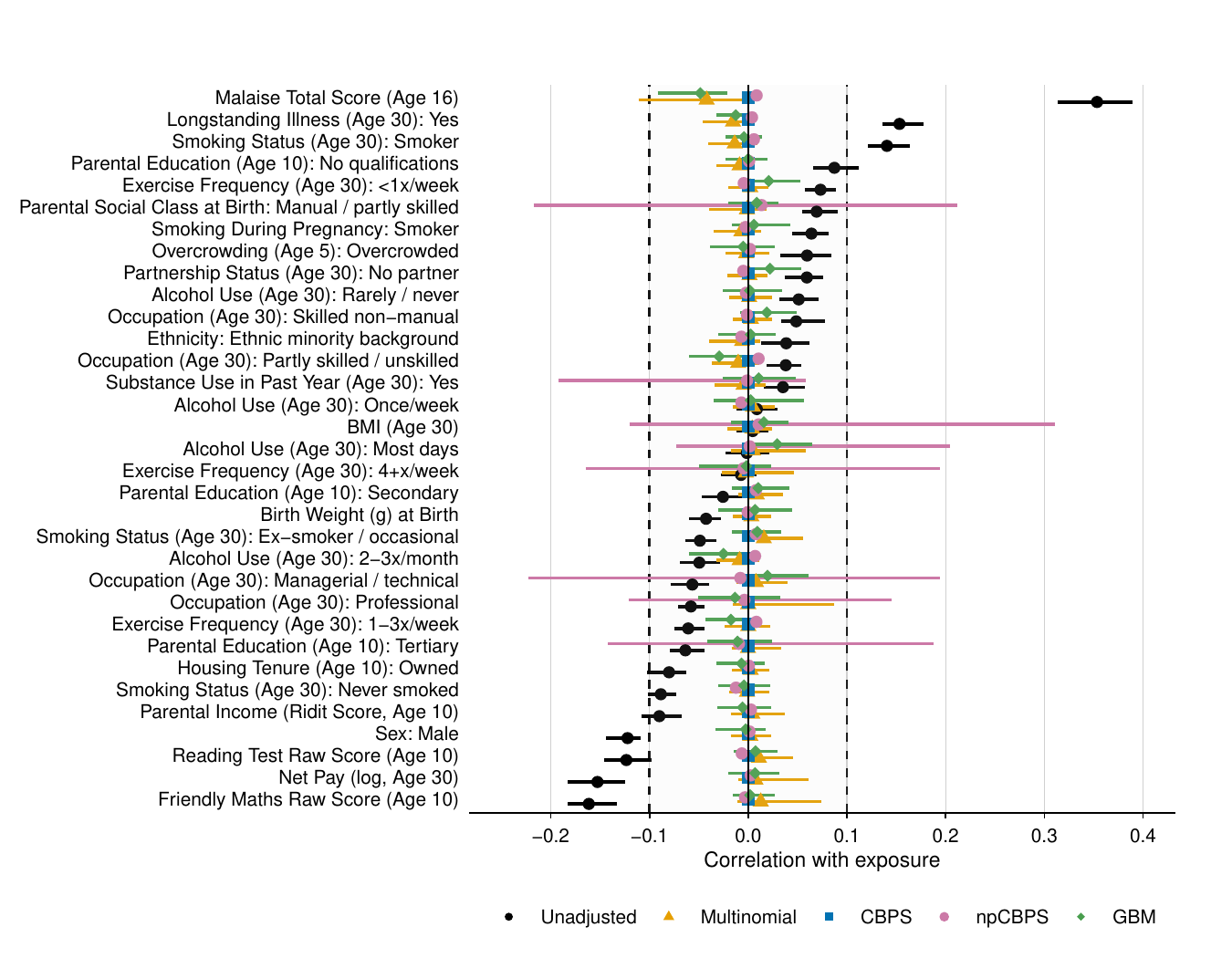}
  \caption{Balance plot for the motivating example including npCBPS alongside other approaches. Points show weighted or unadjusted Pearson and point-biserial correlations between the exposure and each covariate; horizontal whiskers show the range of the correlations across \num[round-precision = 0]{\bcsM} multiply imputed datasets. Data are from the 1970 British Cohort Study (N = \num[round-precision = 0]{\bcsEligible}). (np)CBPS -- (Non-Parametric) Covariate Balancing Propensity Score; GBM -- Generalised Boosted Models.}
  \label{fig:bcs70_balance_npcbps}
\end{figure}

\clearpage

\begin{landscape}

{\footnotesize
\begin{longtable}{>{\raggedright\arraybackslash}p{2.5cm} >{\raggedright\arraybackslash}p{3.5cm} >{\raggedright\arraybackslash}p{2.3cm} >{\raggedright\arraybackslash}p{4.7cm} >{\raggedright\arraybackslash}p{7.5cm}}
\caption{Covariates used in the motivating example from the 1970 British Cohort Study (BCS70).}
\label{tab:bcs70_measures} \\

\toprule
\textbf{Domain} & \textbf{Covariate} & \textbf{Sweep (Age)} & \textbf{Measure(s)} & \textbf{Values} \\
\midrule
\endfirsthead

\multicolumn{5}{l}{\small\textit{Table~\ref{tab:bcs70_measures} continued}} \\[4pt]
\toprule
\textbf{Domain} & \textbf{Covariate} & \textbf{Sweep (Age)} & \textbf{Measure(s)} & \textbf{Values} \\
\midrule
\endhead

\midrule
\multicolumn{5}{r}{\small\textit{Continued on next page}} \\
\endfoot

\bottomrule
\endlastfoot

Demographic & Sex at birth & 1 (birth) & -- & Binary: Female/Male \\
 & Ethnicity & 6 (30) & -- & Binary: White/Non-white \\
 & Partnership & 6 (30) & -- & Binary: Married or partnered/Not partnered \\

\addlinespace[4pt]

Developmental & & & & \\

\quad \textit{Early life SEP} & Social class at birth & 1 (birth) & -- & Binary: Manual/Non-manual \\
 & Overcrowding & 2 (5) & Persons per room (excl.\ kitchen and toilet) & Binary: $\leq$1 / $>$1 people per room \\
 & Combined gross parental income & 3 (10) & -- & Ridit scores \\
 & Household tenure & 3 (10) & -- & Binary: Owned/Not owned \\
 & Combined parental education & 3 (10) & -- & Categorical: No qualifications / Secondary / Tertiary \\

\addlinespace[4pt]

\quad \textit{Mental health} & Psychological distress in adolescence & 4 (16) & Malaise Inventory & Continuous (24-item version) \\

\addlinespace[4pt]

\quad \textit{Other} & Birthweight & 1 (birth) & Birthweight in grams & Continuous \\
 & Smoking in pregnancy & 1 (birth) & -- & Binary: Did not smoke / Smoked \\
 & Early life cognitive ability & 3 (10) & Edinburgh Reading Test and Friendly Maths Test (derived scores) & Continuous (two scores) \\

\addlinespace[4pt]

Socioeconomic & Income & 6 (30) & Weekly net pay (log-transformed) & Continuous \\
 & Occupation & 6 (30) & Occupational class (SOC90) & Categorical: Professional / Managerial-technical / Skilled non-manual / Other \\

\addlinespace[4pt]

Health & Health conditions & 6 (30) & Longstanding illness & Binary \\
 & Exercise & 6 (30) & Frequency of exercise activity & Categorical: $<$1/week / 1--3 times/week / 4+ times/week \\
 & Smoking & 6 (30) & Current smoking status & Categorical: Never / Ex or occasional / Current \\
 & Alcohol consumption & 6 (30) & Frequency of alcoholic drink & Categorical: Most days / 2--3 days/week / Once a week / Less often or never \\
 & Adiposity & 6 (30) & Body mass index & Continuous \\
 & Substance use & 6 (30) & Tried any illegal drug in past 12 months & Binary: Did not try / Tried \\

\end{longtable}

\vspace{-6pt}
\noindent\textit{Note.} Potential confounders were sourced from  sweeps preceding the exposure to ensure temporality and avoid post-treatment biases. SEP -- socioeconomic position.  
}

\end{landscape}

\clearpage

\begin{table}[htbp]
\centering
\begin{threeparttable}
\caption{\label{tab:missingness_parameters}Calibrated missingness parameters used to generate MCAR and MAR scenarios.}
\begin{tabular}[t]{llccccc}
\toprule
\textbf{Mechanism} & \textbf{Exposure DGM} & \textbf{$\phi^\ast$} & \textbf{$C_2$} & \textbf{$C_3$} & \textbf{$A$} & \textbf{$Y$}\\
\midrule
\addlinespace[0.3em]
\multicolumn{7}{l}{\textbf{Per-variable missingness probabilities $p_i$}}\\
MCAR & Both & 0.20 & 0.0213 & 0.0431 & 0.0648 & 0.0866\\
MCAR & Both & 0.40 & 0.0465 & 0.0943 & 0.1422 & 0.1900\\
MCAR & Both & 0.60 & 0.0799 & 0.1598 & 0.2396 & 0.3195\\
MAR & Negative binomial & 0.20 & 0.0234 & 0.0481 & 0.0727 & 0.0974\\
MAR & Negative binomial & 0.40 & 0.0551 & 0.1124 & 0.1697 & 0.2269\\
MAR & Negative binomial & 0.60 & 0.1068 & 0.2016 & 0.2964 & 0.3912\\
MAR & Poisson & 0.20 & 0.0240 & 0.0494 & 0.0748 & 0.1003\\
MAR & Poisson & 0.40 & 0.0570 & 0.1157 & 0.1743 & 0.2329\\
MAR & Poisson & 0.60 & 0.1101 & 0.2064 & 0.3027 & 0.3990\\
\addlinespace[0.3em]
\multicolumn{7}{l}{\textbf{MAR logistic intercepts $\gamma_{0,i}$}}\\
MAR & Negative binomial & 0.20 & -4.3436 & -4.1056 & -4.2076 & -6.9927\\
MAR & Negative binomial & 0.40 & -3.4499 & -3.0772 & -2.9161 & -4.7130\\
MAR & Negative binomial & 0.60 & -2.7200 & -2.2557 & -1.8491 & -2.9763\\
MAR & Poisson & 0.20 & -4.3191 & -4.0741 & -4.1679 & -6.6487\\
MAR & Poisson & 0.40 & -3.4124 & -3.0393 & -2.8697 & -4.7175\\
MAR & Poisson & 0.60 & -2.6849 & -2.2193 & -1.8030 & -3.1232\\
\bottomrule
\end{tabular}
\begin{tablenotes}
\item \textit{Note:}
For MCAR, the same $p_i$ values were used for both exposure DGMs. For MAR, both $p_i$ and intercepts were calibrated separately for each DGM. Values are shown for the four amputed variables in the order used in the simulation: confounders $C_2$, $C_3$, exposure $A$, and outcome $Y$. DGM \textendash{} Data-Generating Mechanism; MAR \textendash{} Missing at Random; MCAR \textendash{} Missing Completely at Random; $\phi^\ast$ \textendash{} target proportion of incomplete rows.
\end{tablenotes}
\end{threeparttable}
\end{table}

\clearpage

\begin{table}[htbp]
\centering
\begin{threeparttable}
\caption{\label{tab:sim_noexpmiss}Relative bias and coverage (MCSE) when the exposure is not subject to missingness.}
\begin{tabular}[t]{lcccc}
\toprule
\multicolumn{2}{c}{\textbf{ }} & \multicolumn{3}{c}{\textbf{MAR}} \\
\cmidrule(l{3pt}r{3pt}){3-5}
\textbf{Method} & \textbf{Complete} & \textbf{20\%} & \textbf{40\%} & \textbf{60\%}\\
\midrule
\addlinespace[0.3em]
\multicolumn{5}{l}{\textbf{Relative bias (\%)}}\\
\hspace{1em}Unadjusted & 26.8 (0.6) & 24.6 (1.0) & 25.9 (1.4) & 28.0 (1.8)\\
\hspace{1em}Adjusted & -1.5 (0.6) & -3.6 (1.0) & -1.9 (1.4) & 0.8 (1.7)\\
\hspace{1em}Multinomial & -1.2 (0.7) & -2.3 (1.1) & -0.6 (1.4) & 2.5 (1.8)\\
\hspace{1em}CBPS & -2.3 (0.7) & -3.3 (1.1) & -1.5 (1.4) & 1.7 (1.8)\\
\addlinespace[0.3em]
\multicolumn{5}{l}{\textbf{Coverage}}\\
\hspace{1em}Unadjusted & 0.823 (0.009) & 0.894 (0.007) & 0.895 (0.007) & 0.904 (0.007)\\
\hspace{1em}Adjusted & 0.958 (0.004) & 0.946 (0.005) & 0.940 (0.005) & 0.932 (0.006)\\
\hspace{1em}Multinomial & 0.952 (0.005) & 0.946 (0.005) & 0.947 (0.005) & 0.939 (0.005)\\
\hspace{1em}CBPS & 0.949 (0.005) & 0.945 (0.005) & 0.949 (0.005) & 0.940 (0.005)\\
\bottomrule
\end{tabular}
\begin{tablenotes}
\item \textit{Note:}
Entries are Monte Carlo summaries over 2,000 simulated datasets (negative binomial exposure DGM, RR = 1.1). Percentages under MAR indicate the percentage of incomplete rows introduced. Missingness was imposed on covariates and the outcome only; the exposure was always fully observed (not imputed). Numbers in brackets indicate Monte Carlo standard errors. The unadjusted and adjusted estimators use unweighted Poisson regression with a sandwich variance estimator. The multinomial and CBPS estimators use inverse probability weighted Poisson regression with a sandwich variance estimator. For each repetition, results were pooled across multiply imputed datasets, with the number of imputations set to the percentage of incomplete cases. CBPS -- Covariate Balancing Propensity Score; MAR -- Missing at random; MCSE -- Monte Carlo standard error.
\end{tablenotes}
\end{threeparttable}
\end{table}

\clearpage

\begingroup\fontsize{8}{10}\selectfont

\begin{ThreePartTable}
\begin{TableNotes}[para]
\item \textit{Note:} 
Descriptive statistics for the eligible sample (N = 16,638). N -- count; SD -- standard deviation.
\end{TableNotes}
\begin{longtable}[c]{lcc}
\captionsetup{justification=centering}
\caption{\label{tab:descriptives}Descriptive statistics for the BCS70 motivating example}\\
\toprule
\textbf{Variable} & \textbf{N (\%) / Mean (SD)} & \textbf{Missing}\\
\midrule
\endfirsthead
\caption[]{Descriptive statistics for the BCS70 motivating example \textit{(continued)}}\\
\toprule
\textbf{Variable} & \textbf{N (\%) / Mean (SD)} & \textbf{Missing}\\
\midrule
\endhead

\endfoot
\bottomrule
\insertTableNotes
\endlastfoot
\addlinespace[0.3em]
\multicolumn{3}{l}{\textbf{Demographics}}\\
\hspace{1em}Female & 8,107 (48.7\%) & 0 (0\%)\\
\hspace{1em}Country at Birth &  & 26 (0.2\%)\\
\hspace{1em}\hspace{1em}England & 13,804 (83.1\%) & \\
\hspace{1em}\hspace{1em}Scotland & 1,567 (9.4\%) & \\
\hspace{1em}\hspace{1em}Wales & 833 (5.0\%) & \\
\hspace{1em}\hspace{1em}Other & 408 (2.5\%) & \\
\hspace{1em}Ethnicity: Non-White & 405 (3.7\%) & 5,635 (33.9\%)\\
\hspace{1em}Married/Partnered (Age 30) & 7,285 (66.6\%) & 5,699 (34.3\%)\\
\addlinespace[0.3em]
\multicolumn{3}{l}{\textbf{Socioeconomic Variables}}\\
\hspace{1em}Parental Social Class at Birth &  & 1,379 (8.3\%)\\
\hspace{1em}\hspace{1em}Non-Manual & 4,824 (31.6\%) & \\
\hspace{1em}\hspace{1em}Manual/Unskilled & 10,338 (67.8\%) & \\
\hspace{1em}\hspace{1em}Other & 97 (0.6\%) & \\
\hspace{1em}Overcrowded (Age 5) & 2,231 (18.1\%) & 4,317 (25.9\%)\\
\hspace{1em}Housing tenure: Owned (Age 10) & 7,916 (61.1\%) & 3,682 (22.1\%)\\
\hspace{1em}Parental Education (Age 10) &  & 5,328 (32.0\%)\\
\hspace{1em}\hspace{1em}No Qualifications & 3,534 (31.2\%) & \\
\hspace{1em}\hspace{1em}Secondary & 4,315 (38.2\%) & \\
\hspace{1em}\hspace{1em}Tertiary & 3,461 (30.6\%) & \\
\hspace{1em}Parental Income (Age 10) &  & 4,689 (28.2\%)\\
\hspace{1em}\hspace{1em}<£50 & 866 (7.2\%) & \\
\hspace{1em}\hspace{1em}£50-99 & 3,631 (30.4\%) & \\
\hspace{1em}\hspace{1em}£100-149 & 4,074 (34.1\%) & \\
\hspace{1em}\hspace{1em}£150-199 & 1,944 (16.3\%) & \\
\hspace{1em}\hspace{1em}£200+ & 1,434 (12.0\%) & \\
\hspace{1em}Occupation (Age 30) &  & 7,670 (46.1\%)\\
\hspace{1em}\hspace{1em}Partly Skilled/Unskilled/Manual & 3,082 (34.4\%) & \\
\hspace{1em}\hspace{1em}Skilled Non-Manual & 2,215 (24.7\%) & \\
\hspace{1em}\hspace{1em}Managerial/Technical & 3,109 (34.7\%) & \\
\hspace{1em}\hspace{1em}Professional & 562 (6.3\%) & \\
\addlinespace[0.3em]
\multicolumn{3}{l}{\textbf{Health Variables}}\\
\hspace{1em}Birth Weight (g) at Birth & 3,302.6 (526.4) & 1,375 (8.3\%)\\
\hspace{1em}Malaise Total Score (Age 16) & 9.2 (5.5) & 11,436 (68.7\%)\\
\hspace{1em}BMI (Age 30) & 24.9 (4.5) & 5,946 (35.7\%)\\
\hspace{1em}Longstanding Illness (Age 30) & 2,530 (23.0\%) & 5,651 (34.0\%)\\
\hspace{1em}Malaise Score (Age 34) & 1.7 (1.9) & 7,152 (43.0\%)\\
\hspace{1em}Longstanding Illness (Age 42) & 2,845 (29.1\%) & 6,876 (41.3\%)\\
\addlinespace[0.3em]
\multicolumn{3}{l}{\textbf{Health Behaviours}}\\
\hspace{1em}Smoking During Pregnancy & 6,236 (41.0\%) & 1,441 (8.7\%)\\
\hspace{1em}Exercise Frequency (Age 30) &  & 5,656 (34.0\%)\\
\hspace{1em}\hspace{1em}<1 Time/Week & 3,198 (29.1\%) & \\
\hspace{1em}\hspace{1em}1-3 Times/Week & 4,740 (43.2\%) & \\
\hspace{1em}\hspace{1em}4+ Times/Week & 3,044 (27.7\%) & \\
\hspace{1em}Smoking Status (Age 30) &  & 5,654 (34.0\%)\\
\hspace{1em}\hspace{1em}Smoker & 3,210 (29.2\%) & \\
\hspace{1em}\hspace{1em}Never Smoked & 4,862 (44.3\%) & \\
\hspace{1em}\hspace{1em}Ex-Smoker/Occasional & 2,912 (26.5\%) & \\
\hspace{1em}Alcohol Use (Age 30) &  & 5,654 (34.0\%)\\
\hspace{1em}\hspace{1em}Rarely/Never & 3,638 (33.1\%) & \\
\hspace{1em}\hspace{1em}Once/Week & 2,386 (21.7\%) & \\
\hspace{1em}\hspace{1em}2-3 Times/Month & 3,560 (32.4\%) & \\
\hspace{1em}\hspace{1em}Most Days & 1,400 (12.7\%) & \\
\hspace{1em}Substance Use in Past Year (Age 30) & 1,025 (9.4\%) & 5,750 (34.6\%)\\
\addlinespace[0.3em]
\multicolumn{3}{l}{\textbf{Cognition}}\\
\hspace{1em}Friendly Maths Raw Score (Age 10) & 43.9 (12.3) & 5,538 (33.3\%)\\
\hspace{1em}Reading Test Raw Score (Age 10) & 40.2 (12.7) & 5,529 (33.2\%)\\*
\end{longtable}
\end{ThreePartTable}
\endgroup{}

\bibliographystyle{unsrtnat}
\bibliography{references}

\begin{thebibliography}{55}
\providecommand{\natexlab}[1]{#1}
\providecommand{\url}[1]{\texttt{#1}}
\expandafter\ifx\csname urlstyle\endcsname\relax
  \providecommand{\doi}[1]{doi: #1}\else
  \providecommand{\doi}{doi: \begingroup \urlstyle{rm}\Url}\fi

\bibitem[VanderWeele et~al.(2020)VanderWeele, Mathur, and
  Chen]{vanderweele_outcome-wide_2020}
Tyler~J. VanderWeele, Maya~B. Mathur, and Ying Chen.
\newblock Outcome-{Wide} {Longitudinal} {Designs} for {Causal} {Inference}: {A}
  {New} {Template} for {Empirical} {Studies}.
\newblock \emph{Statistical Science}, 35\penalty0 (3), August 2020.
\newblock ISSN 0883-4237.
\newblock \doi{10.1214/19-STS728}.
\newblock URL
  \url{https://projecteuclid.org/journals/statistical-science/volume-35/issue-3/Outcome-Wide-Longitudinal-Designs-for-Causal-Inference--A-New/10.1214/19-STS728.full}.

\bibitem[Rosenbaum and Rubin(1983)]{rosenbaum_central_1983}
Paul~R. Rosenbaum and Donald~B. Rubin.
\newblock The central role of the propensity score in observational studies for
  causal effects.
\newblock \emph{Biometrika}, 70\penalty0 (1):\penalty0 41--55, April 1983.
\newblock ISSN 0006-3444.
\newblock \doi{10.1093/biomet/70.1.41}.
\newblock URL \url{https://doi.org/10.1093/biomet/70.1.41}.

\bibitem[Robins et~al.(2000)Robins, Hernán, and
  Brumback]{robins_marginal_2000}
James Robins, Miguel~Ángel Hernán, and Babette Brumback.
\newblock Marginal {Structural} {Models} and {Causal} {Inference} in
  {Epidemiology}.
\newblock \emph{Epidemiology}, 11\penalty0 (5):\penalty0 550, September 2000.
\newblock ISSN 1044-3983.
\newblock URL
  \url{https://journals.lww.com/epidem/fulltext/2000/09000/marginal_structural_models_and_causal_inference_in.11.aspx}.

\bibitem[Hirano and Imbens(2004)]{gelman_propensity_2004}
Keisuke Hirano and Guido~W. Imbens.
\newblock The {Propensity} {Score} with {Continuous} {Treatments}.
\newblock In Andrew Gelman and Xiao‐Li Meng, editors, \emph{Wiley {Series} in
  {Probability} and {Statistics}}, pages 73--84. Wiley, 1 edition, July 2004.
\newblock ISBN 978-0-470-09043-5 978-0-470-09045-9.
\newblock URL \url{https://onlinelibrary.wiley.com/doi/10.1002/0470090456.ch7}.

\bibitem[Naimi et~al.(2014)Naimi, Moodie, Auger, and
  Kaufman]{naimi_constructing_2014}
Ashley~I. Naimi, Erica E.~M. Moodie, Nathalie Auger, and Jay~S. Kaufman.
\newblock Constructing {Inverse} {Probability} {Weights} for {Continuous}
  {Exposures}: {A} {Comparison} of {Methods}.
\newblock \emph{Epidemiology}, 25\penalty0 (2):\penalty0 292, March 2014.
\newblock ISSN 1044-3983.
\newblock \doi{10.1097/EDE.0000000000000053}.
\newblock URL
  \url{https://journals.lww.com/epidem/fulltext/2014/03000/constructing_inverse_probability_weights_for.21.aspx}.

\bibitem[Fong et~al.(2018)Fong, Hazlett, and Imai]{fong_covariate_2018}
Christian Fong, Chad Hazlett, and Kosuke Imai.
\newblock Covariate balancing propensity score for a continuous treatment:
  {Application} to the efficacy of political advertisements.
\newblock \emph{The Annals of Applied Statistics}, 12\penalty0 (1), March 2018.
\newblock ISSN 1932-6157.
\newblock \doi{10.1214/17-AOAS1101}.
\newblock URL
  \url{https://projecteuclid.org/journals/annals-of-applied-statistics/volume-12/issue-1/Covariate-balancing-propensity-score-for-a-continuous-treatment--Application/10.1214/17-AOAS1101.full}.

\bibitem[Imai and Ratkovic(2014)]{imai_covariate_2014}
Kosuke Imai and Marc Ratkovic.
\newblock Covariate {Balancing} {Propensity} {Score}.
\newblock \emph{Journal of the Royal Statistical Society Series B: Statistical
  Methodology}, 76\penalty0 (1):\penalty0 243--263, January 2014.
\newblock ISSN 1369-7412, 1467-9868.
\newblock \doi{10.1111/rssb.12027}.
\newblock URL \url{https://academic.oup.com/jrsssb/article/76/1/243/7075938}.

\bibitem[Zhu et~al.(2015)Zhu, Coffman, and Ghosh]{zhu_boosting_2015}
Yeying Zhu, Donna~L. Coffman, and Debashis Ghosh.
\newblock A {Boosting} {Algorithm} for {Estimating} {Generalized} {Propensity}
  {Scores} with {Continuous} {Treatments}.
\newblock \emph{Journal of Causal Inference}, 3\penalty0 (1):\penalty0 25--40,
  March 2015.
\newblock ISSN 2193-3677, 2193-3685.
\newblock \doi{10.1515/jci-2014-0022}.
\newblock URL
  \url{https://www.degruyter.com/document/doi/10.1515/jci-2014-0022/html}.

\bibitem[Székely and Rizzo(2004)]{szekely_testing_2004}
G.~Székely and M.~Rizzo.
\newblock Testing for equal distributions in high dimension.
\newblock \emph{InterStat}, 5\penalty0 (1–6):\penalty0 1249--72, 2004.

\bibitem[Székely and Rizzo(2013)]{szekely_energy_2013}
Gábor~J. Székely and Maria~L. Rizzo.
\newblock Energy statistics: {A} class of statistics based on distances.
\newblock \emph{Journal of Statistical Planning and Inference}, 143\penalty0
  (8):\penalty0 1249--1272, August 2013.
\newblock ISSN 0378-3758.
\newblock \doi{10.1016/j.jspi.2013.03.018}.
\newblock URL
  \url{https://www.sciencedirect.com/science/article/pii/S0378375813000633}.

\bibitem[Huling et~al.(2024)Huling, Greifer, and
  Chen]{huling_independence_2024}
Jared~D. Huling, Noah Greifer, and Guanhua Chen.
\newblock Independence {Weights} for {Causal} {Inference} with {Continuous}
  {Treatments}.
\newblock \emph{Journal of the American Statistical Association}, 119\penalty0
  (546):\penalty0 1657--1670, April 2024.
\newblock ISSN 0162-1459, 1537-274X.
\newblock \doi{10.1080/01621459.2023.2213485}.
\newblock URL
  \url{https://www.tandfonline.com/doi/full/10.1080/01621459.2023.2213485}.

\bibitem[Sack et~al.(2023)Sack, Shepherd, Audet, De~Schacht, and
  Samuels]{sack_inverse_2023}
Daniel~E Sack, Bryan~E Shepherd, Carolyn~M Audet, Caroline De~Schacht, and
  Lauren~R Samuels.
\newblock Inverse {Probability} {Weights} for {Quasicontinuous} {Ordinal}
  {Exposures} {With} a {Binary} {Outcome}: {Method} {Comparison} and {Case}
  {Study}.
\newblock \emph{American Journal of Epidemiology}, 192\penalty0 (7):\penalty0
  1192--1206, July 2023.
\newblock ISSN 0002-9262.
\newblock \doi{10.1093/aje/kwad085}.
\newblock URL \url{https://doi.org/10.1093/aje/kwad085}.

\bibitem[Qu and Lipkovich(2009)]{qu_propensity_2009}
Yongming Qu and Ilya Lipkovich.
\newblock Propensity score estimation with missing values using a multiple
  imputation missingness pattern ({MIMP}) approach.
\newblock \emph{Statistics in Medicine}, 28\penalty0 (9):\penalty0 1402--1414,
  2009.
\newblock ISSN 1097-0258.
\newblock \doi{10.1002/sim.3549}.
\newblock URL \url{https://onlinelibrary.wiley.com/doi/abs/10.1002/sim.3549}.

\bibitem[Seaman and White(2014)]{seaman_inverse_2014}
Shaun~R. Seaman and Ian~R. White.
\newblock Inverse {Probability} {Weighting} with {Missing} {Predictors} of
  {Treatment} {Assignment} or {Missingness}.
\newblock \emph{Communications in Statistics - Theory and Methods}, 43\penalty0
  (16):\penalty0 3499--3515, August 2014.
\newblock ISSN 0361-0926.
\newblock \doi{10.1080/03610926.2012.700371}.
\newblock URL \url{https://doi.org/10.1080/03610926.2012.700371}.

\bibitem[Leyrat et~al.(2019)Leyrat, Seaman, White, Douglas, Smeeth, Kim,
  Resche-Rigon, Carpenter, and Williamson]{leyrat_propensity_2019}
Clémence Leyrat, Shaun~R Seaman, Ian~R White, Ian Douglas, Liam Smeeth, Joseph
  Kim, Matthieu Resche-Rigon, James~R Carpenter, and Elizabeth~J Williamson.
\newblock Propensity score analysis with partially observed covariates: {How}
  should multiple imputation be used?
\newblock \emph{Statistical Methods in Medical Research}, 28\penalty0
  (1):\penalty0 3--19, January 2019.
\newblock ISSN 0962-2802.
\newblock \doi{10.1177/0962280217713032}.
\newblock URL \url{https://doi.org/10.1177/0962280217713032}.
\newblock Publisher: SAGE Publications Ltd STM.

\bibitem[Nguyen and Stuart(2024)]{nguyen_multiple_2024}
Trang~Quynh Nguyen and Elizabeth~A Stuart.
\newblock Multiple imputation for propensity score analysis with covariates
  missing at random: some clarity on “ \textit{within} ” and “
  \textit{across} ” methods.
\newblock \emph{American Journal of Epidemiology}, 193\penalty0 (10):\penalty0
  1470--1476, October 2024.
\newblock ISSN 0002-9262, 1476-6256.
\newblock \doi{10.1093/aje/kwae105}.
\newblock URL \url{https://academic.oup.com/aje/article/193/10/1470/7691217}.

\bibitem[Morris et~al.(2019)Morris, White, and Crowther]{morris_using_2019}
Tim~P. Morris, Ian~R. White, and Michael~J. Crowther.
\newblock Using simulation studies to evaluate statistical methods.
\newblock \emph{Statistics in Medicine}, 38\penalty0 (11):\penalty0 2074--2102,
  2019.
\newblock ISSN 1097-0258.
\newblock \doi{10.1002/sim.8086}.
\newblock URL \url{https://onlinelibrary.wiley.com/doi/abs/10.1002/sim.8086}.

\bibitem[Imbens(2000)]{imbens_role_2000}
G.~Imbens.
\newblock The role of the propensity score in estimating dose-response
  functions.
\newblock \emph{Biometrika}, 87\penalty0 (3):\penalty0 706--710, September
  2000.
\newblock ISSN 0006-3444, 1464-3510.
\newblock \doi{10.1093/biomet/87.3.706}.
\newblock URL
  \url{https://academic.oup.com/biomet/article-lookup/doi/10.1093/biomet/87.3.706}.

\bibitem[Cole and Hernán(2008)]{cole_constructing_2008}
Stephen~R. Cole and Miguel~A. Hernán.
\newblock Constructing {Inverse} {Probability} {Weights} for {Marginal}
  {Structural} {Models}.
\newblock \emph{American Journal of Epidemiology}, 168\penalty0 (6):\penalty0
  656--664, September 2008.
\newblock ISSN 0002-9262.
\newblock \doi{10.1093/aje/kwn164}.
\newblock URL \url{https://doi.org/10.1093/aje/kwn164}.

\bibitem[Goetghebeur et~al.(2020)Goetghebeur, Le~Cessie, De~Stavola, Moodie,
  Waernbaum, and {“on behalf of” the topic group Causal Inference (TG7) of
  the STRATOS initiative}]{goetghebeur_formulating_2020}
Els Goetghebeur, Saskia Le~Cessie, Bianca De~Stavola, Erica~Em Moodie, Ingeborg
  Waernbaum, and {“on behalf of” the topic group Causal Inference (TG7) of
  the STRATOS initiative}.
\newblock Formulating causal questions and principled statistical answers.
\newblock \emph{Statistics in Medicine}, 39\penalty0 (30):\penalty0 4922--4948,
  December 2020.
\newblock ISSN 0277-6715, 1097-0258.
\newblock \doi{10.1002/sim.8741}.
\newblock URL \url{https://onlinelibrary.wiley.com/doi/10.1002/sim.8741}.

\bibitem[Colnet et~al.(2025)Colnet, Josse, Varoquaux, and
  Scornet]{colnet_risk_2025}
Bénédicte Colnet, Julie Josse, Gaël Varoquaux, and Erwan Scornet.
\newblock Risk ratio, odds ratio, risk difference... {Which} causal measure is
  easier to generalize?, September 2025.
\newblock URL \url{http://arxiv.org/abs/2303.16008}.
\newblock arXiv:2303.16008.

\bibitem[Friedman(2001)]{friedman_greedy_2001}
Jerome~H. Friedman.
\newblock Greedy function approximation: {A} gradient boosting machine.
\newblock \emph{The Annals of Statistics}, 29\penalty0 (5), October 2001.
\newblock ISSN 0090-5364.
\newblock \doi{10.1214/aos/1013203451}.
\newblock URL
  \url{https://projecteuclid.org/journals/annals-of-statistics/volume-29/issue-5/Greedy-function-approximation-A-gradient-boosting-machine/10.1214/aos/1013203451.full}.

\bibitem[McCaffrey et~al.(2004)McCaffrey, Ridgeway, and
  Morral]{mccaffrey_propensity_2004}
Daniel~F. McCaffrey, Greg Ridgeway, and Andrew~R. Morral.
\newblock Propensity {Score} {Estimation} {With} {Boosted} {Regression} for
  {Evaluating} {Causal} {Effects} in {Observational} {Studies}.
\newblock \emph{Psychological Methods}, 9\penalty0 (4):\penalty0 403--425,
  December 2004.
\newblock ISSN 1939-1463, 1082-989X.
\newblock \doi{10.1037/1082-989X.9.4.403}.
\newblock URL \url{https://doi.apa.org/doi/10.1037/1082-989X.9.4.403}.

\bibitem[McCaffrey et~al.(2013)McCaffrey, Griffin, Almirall, Slaughter,
  Ramchand, and Burgette]{mccaffrey_tutorial_2013}
Daniel~F. McCaffrey, Beth~Ann Griffin, Daniel Almirall, Mary~Ellen Slaughter,
  Rajeev Ramchand, and Lane~F. Burgette.
\newblock A tutorial on propensity score estimation for multiple treatments
  using generalized boosted models.
\newblock \emph{Statistics in Medicine}, 32\penalty0 (19):\penalty0 3388--3414,
  August 2013.
\newblock ISSN 0277-6715, 1097-0258.
\newblock \doi{10.1002/sim.5753}.
\newblock URL \url{https://onlinelibrary.wiley.com/doi/10.1002/sim.5753}.

\bibitem[Huling and Mak(2024)]{huling_energy_2024}
Jared~D. Huling and Simon Mak.
\newblock Energy balancing of covariate distributions.
\newblock \emph{Journal of Causal Inference}, 12\penalty0 (1), January 2024.
\newblock ISSN 2193-3685.
\newblock \doi{10.1515/jci-2022-0029}.
\newblock URL
  \url{https://www.degruyter.com/document/doi/10.1515/jci-2022-0029/html?srsltid=AfmBOooa090Yir8f5ayYoezOP5wD9xzpM-mtZTboXbbJxxLa1fNLhy7X}.

\bibitem[Elliott and Shepherd(2006)]{elliott_cohort_2006}
Jane Elliott and Peter Shepherd.
\newblock Cohort {Profile}: 1970 {British} {Birth} {Cohort} ({BCS70}).
\newblock \emph{International Journal of Epidemiology}, 35\penalty0
  (4):\penalty0 836--843, August 2006.
\newblock ISSN 1464-3685, 0300-5771.
\newblock \doi{10.1093/ije/dyl174}.
\newblock URL
  \url{http://academic.oup.com/ije/article/35/4/836/686544/Cohort-Profile-1970-British-Birth-Cohort-BCS70}.

\bibitem[Sullivan et~al.(2023)Sullivan, Brown, Hamer, and
  Ploubidis]{sullivan_cohort_2023}
Alice Sullivan, Matt Brown, Mark Hamer, and George~B Ploubidis.
\newblock Cohort {Profile} {Update}: {The} 1970 {British} {Cohort} {Study}
  ({BCS70}).
\newblock \emph{International Journal of Epidemiology}, 52\penalty0
  (3):\penalty0 e179--e186, June 2023.
\newblock ISSN 0300-5771, 1464-3685.
\newblock \doi{10.1093/ije/dyac148}.
\newblock URL \url{https://academic.oup.com/ije/article/52/3/e179/6645761}.

\bibitem[McElroy et~al.(2020)McElroy, Villadsen, Patalay, Goodman, Richards,
  Northstone, Fearon, Tibber, Gondek, and
  Ploubidis]{mcelroy_harmonisation_2020}
Eoin McElroy, Aase Villadsen, Praveetha Patalay, Alissa Goodman, Marcus
  Richards, Kate Northstone, Pasco Fearon, Marc Tibber, Dawid Gondek, and
  George~B. Ploubidis.
\newblock Harmonisation and measurement properties of mental health measures in
  six {British} cohorts.
\newblock Technical report, CLOSER, London, UK, 2020.

\bibitem[Ploubidis et~al.(2019)Ploubidis, McElroy, and
  Moreira]{ploubidis_longitudinal_2019}
George~B. Ploubidis, Eoin McElroy, and Hugo~Cogo Moreira.
\newblock A longitudinal examination of the measurement equivalence of mental
  health assessments in two {British} birth cohorts.
\newblock \emph{Longitudinal and Life Course Studies}, 10\penalty0
  (4):\penalty0 471--489, October 2019.
\newblock ISSN 1757-9597.
\newblock \doi{10.1332/175795919X15683588979486}.
\newblock URL
  \url{https://bristoluniversitypressdigital.com/view/journals/llcs/10/4/article-p471.xml}.

\bibitem[for Health Improvement
  \&~Disparities(2024)]{office_for_health_improvement__disparities_cardiovascular_2024}
Office for Health Improvement \&~Disparities.
\newblock Cardiovascular disease and diabetes profiles: statistical commentary.
\newblock Technical report, Office for Health Improvement \& Disparities, 2024.
\newblock URL
  \url{https://www.gov.uk/government/statistics/cardiovascular-disease-and-diabetes-profiles-march-2024-update/cardiovascular-disease-and-diabetes-profiles-statistical-commentary}.

\bibitem[Oberman and Vink(2024)]{oberman_toward_2024}
Hanne~I. Oberman and Gerko Vink.
\newblock Toward a standardized evaluation of imputation methodology.
\newblock \emph{Biometrical Journal}, 66\penalty0 (1):\penalty0 2200107, 2024.
\newblock ISSN 1521-4036.
\newblock \doi{10.1002/bimj.202200107}.
\newblock URL
  \url{https://onlinelibrary.wiley.com/doi/abs/10.1002/bimj.202200107}.

\bibitem[Morris et~al.(2024)Morris, White, Cro, Bartlett, Carpenter, and
  Pham]{morris_comment_2024}
Tim~P. Morris, Ian~R. White, Suzie Cro, Jonathan~W. Bartlett, James~R.
  Carpenter, and Tra~My Pham.
\newblock Comment on {Oberman} \& {Vink}: {Should} we fix or simulate the
  complete data in simulation studies evaluating missing data methods?
\newblock \emph{Biometrical Journal}, 66\penalty0 (1):\penalty0 2300085,
  January 2024.
\newblock ISSN 0323-3847, 1521-4036.
\newblock \doi{10.1002/bimj.202300085}.
\newblock URL \url{https://onlinelibrary.wiley.com/doi/10.1002/bimj.202300085}.

\bibitem[Zou(2004)]{zou_modified_2004}
Guangyong Zou.
\newblock A {Modified} {Poisson} {Regression} {Approach} to {Prospective}
  {Studies} with {Binary} {Data}.
\newblock \emph{American Journal of Epidemiology}, 159\penalty0 (7):\penalty0
  702--706, April 2004.
\newblock ISSN 0002-9262.
\newblock \doi{10.1093/aje/kwh090}.
\newblock URL \url{https://doi.org/10.1093/aje/kwh090}.

\bibitem[Greifer(2024)]{greifer_weightit_2024}
Noah Greifer.
\newblock {WeightIt}: {Weighting} for {Covariate} {Balance} in {Observational}
  {Studies}.
\newblock Technical report, 2024.
\newblock URL \url{https://ngreifer.github.io/WeightIt/}.
\newblock R package version 1.3.0.

\bibitem[van Buuren and Groothuis-Oudshoorn(2011)]{van_buuren_mice_2011}
Stef van Buuren and Karin Groothuis-Oudshoorn.
\newblock \textbf{mice} : {Multivariate} {Imputation} by {Chained} {Equations}
  in \textit{{R}}.
\newblock \emph{Journal of Statistical Software}, 45\penalty0 (3), 2011.
\newblock ISSN 1548-7660.
\newblock \doi{10.18637/jss.v045.i03}.
\newblock URL \url{http://www.jstatsoft.org/v45/i03/}.

\bibitem[White et~al.(2011)White, Royston, and Wood]{white_multiple_2011}
Ian~R. White, Patrick Royston, and Angela~M. Wood.
\newblock Multiple imputation using chained equations: {Issues} and guidance
  for practice.
\newblock \emph{Statistics in Medicine}, 30\penalty0 (4):\penalty0 377--399,
  February 2011.
\newblock ISSN 0277-6715, 1097-0258.
\newblock \doi{10.1002/sim.4067}.
\newblock URL \url{https://onlinelibrary.wiley.com/doi/10.1002/sim.4067}.

\bibitem[Rubin(2004)]{rubin_multiple_2004}
Donald~B. Rubin.
\newblock \emph{Multiple imputation for nonresponse in surveys}.
\newblock Wiley, 2004.
\newblock ISBN 978-0-471-65574-9.

\bibitem[Kong et~al.(1994)Kong, Liu, and Wong]{kong_sequential_1994}
Augustine Kong, Jun~S. Liu, and Wing~Hung Wong.
\newblock Sequential {Imputations} and {Bayesian} {Missing} {Data} {Problems}.
\newblock \emph{Journal of the American Statistical Association}, 89\penalty0
  (425):\penalty0 278--288, March 1994.
\newblock ISSN 0162-1459.
\newblock \doi{10.1080/01621459.1994.10476469}.
\newblock URL
  \url{https://www.tandfonline.com/doi/abs/10.1080/01621459.1994.10476469}.

\bibitem[Austin(2019)]{austin_assessing_2019}
Peter~C Austin.
\newblock Assessing covariate balance when using the generalized propensity
  score with quantitative or continuous exposures.
\newblock \emph{Statistical Methods in Medical Research}, 28\penalty0
  (5):\penalty0 1365--1377, May 2019.
\newblock ISSN 0962-2802, 1477-0334.
\newblock \doi{10.1177/0962280218756159}.
\newblock URL \url{https://journals.sagepub.com/doi/10.1177/0962280218756159}.

\bibitem[Arias-de La~Torre et~al.(2021)Arias-de La~Torre, Ronaldson, Prina,
  Matcham, Pinto~Pereira, Hatch, Armstrong, Pickles, Hotopf, and
  Dregan]{arias-de_la_torre_depressive_2021}
Jorge Arias-de La~Torre, Amy Ronaldson, Matthew Prina, Faith Matcham, Snehal~M
  Pinto~Pereira, Stephani~L Hatch, David Armstrong, Andrew Pickles, Matthew
  Hotopf, and Alex Dregan.
\newblock Depressive symptoms during early adulthood and the development of
  physical multimorbidity in the {UK}: an observational cohort study.
\newblock \emph{The Lancet Healthy Longevity}, 2\penalty0 (12):\penalty0
  e801--e810, December 2021.
\newblock ISSN 26667568.
\newblock \doi{10.1016/S2666-7568(21)00259-2}.
\newblock URL
  \url{https://linkinghub.elsevier.com/retrieve/pii/S2666756821002592}.

\bibitem[Gondek et~al.(2022)Gondek, Bann, Patalay, Goodman, McElroy, Richards,
  and Ploubidis]{gondek_psychological_2022}
Dawid Gondek, David Bann, Praveetha Patalay, Alissa Goodman, Eoin McElroy,
  Marcus Richards, and George~B. Ploubidis.
\newblock Psychological distress from early adulthood to early old age:
  evidence from the 1946, 1958 and 1970 {British} birth cohorts.
\newblock \emph{Psychological Medicine}, 52\penalty0 (8):\penalty0 1471--1480,
  June 2022.
\newblock ISSN 0033-2917, 1469-8978.
\newblock \doi{10.1017/S003329172000327X}.
\newblock URL
  \url{https://www.cambridge.org/core/product/identifier/S003329172000327X/type/journal_article}.

\bibitem[Katsoulis et~al.(2024)Katsoulis, Narayanan, Dodgeon, Ploubidis, and
  Silverwood]{katsoulis_data_2024}
Michail Katsoulis, Martina Narayanan, Brian Dodgeon, George Ploubidis, and
  Richard Silverwood.
\newblock A data driven approach to address missing data in the 1970 {British}
  birth cohort, February 2024.
\newblock URL \url{http://medrxiv.org/lookup/doi/10.1101/2024.02.01.24302101}.

\bibitem[Griffin et~al.(2017)Griffin, McCaffrey, Almirall, Burgette, and
  Setodji]{griffin_chasing_2017}
Beth~Ann Griffin, Daniel~F. McCaffrey, Daniel Almirall, Lane~F. Burgette, and
  Claude~Messan Setodji.
\newblock Chasing {Balance} and {Other} {Recommendations} for {Improving}
  {Nonparametric} {Propensity} {Score} {Models}.
\newblock \emph{Journal of Causal Inference}, 5\penalty0 (2), September 2017.
\newblock ISSN 2193-3685.
\newblock \doi{10.1515/jci-2015-0026}.
\newblock URL
  \url{https://www.degruyterbrill.com/document/doi/10.1515/jci-2015-0026/html}.

\bibitem[Lee et~al.(2011)Lee, Lessler, and Stuart]{lee_weight_2011}
Brian~K. Lee, Justin Lessler, and Elizabeth~A. Stuart.
\newblock Weight {Trimming} and {Propensity} {Score} {Weighting}.
\newblock \emph{PLoS ONE}, 6\penalty0 (3):\penalty0 e18174, March 2011.
\newblock ISSN 1932-6203.
\newblock \doi{10.1371/journal.pone.0018174}.
\newblock URL \url{https://dx.plos.org/10.1371/journal.pone.0018174}.

\bibitem[von Hippel(2025)]{von_hippel_imputing_2025}
Paul~T. von Hippel.
\newblock Imputing {With} {Predictive} {Mean} {Matching} {Can} {Be} {Severely}
  {Biased} {When} {Values} {Are} {Missing} {At} {Random}, 2025.
\newblock URL \url{https://arxiv.org/abs/2506.22981}.

\bibitem[Kleinke and Reinecke(2013)]{kleinke_multiple_2013}
Kristian Kleinke and Jost Reinecke.
\newblock Multiple imputation of incomplete zero‐inflated count data.
\newblock \emph{Statistica Neerlandica}, 67\penalty0 (3):\penalty0 311--336,
  August 2013.
\newblock ISSN 0039-0402, 1467-9574.
\newblock \doi{10.1111/stan.12009}.
\newblock URL \url{https://onlinelibrary.wiley.com/doi/10.1111/stan.12009}.

\bibitem[Setodji et~al.(2017)Setodji, McCaffrey, Burgette, Almirall, and
  Griffin]{setodji_right_2017}
Claude~M. Setodji, Daniel~F. McCaffrey, Lane~F. Burgette, Daniel Almirall, and
  Beth~Ann Griffin.
\newblock The {Right} {Tool} for the {Job}: {Choosing} {Between}
  {Covariate}-balancing and {Generalized} {Boosted} {Model} {Propensity}
  {Scores}.
\newblock \emph{Epidemiology}, 28\penalty0 (6):\penalty0 802--811, November
  2017.
\newblock ISSN 1044-3983.
\newblock \doi{10.1097/EDE.0000000000000734}.
\newblock URL \url{http://journals.lww.com/00001648-201711000-00007}.

\bibitem[Petersen et~al.(2012)Petersen, Porter, Gruber, Wang, and Van
  Der~Laan]{petersen_diagnosing_2012}
Maya~L Petersen, Kristin~E Porter, Susan Gruber, Yue Wang, and Mark~J Van
  Der~Laan.
\newblock Diagnosing and responding to violations in the positivity assumption.
\newblock \emph{Statistical Methods in Medical Research}, 21\penalty0
  (1):\penalty0 31--54, February 2012.
\newblock ISSN 0962-2802, 1477-0334.
\newblock \doi{10.1177/0962280210386207}.
\newblock URL \url{https://journals.sagepub.com/doi/10.1177/0962280210386207}.

\bibitem[Zhou et~al.(2020)Zhou, Matsouaka, and Thomas]{zhou_propensity_2020}
Yunji Zhou, Roland~A Matsouaka, and Laine Thomas.
\newblock Propensity score weighting under limited overlap and model
  misspecification.
\newblock \emph{Statistical Methods in Medical Research}, 29\penalty0
  (12):\penalty0 3721--3756, December 2020.
\newblock ISSN 0962-2802, 1477-0334.
\newblock \doi{10.1177/0962280220940334}.
\newblock URL \url{https://journals.sagepub.com/doi/10.1177/0962280220940334}.

\bibitem[Yiu and Su(2018)]{yiu_covariate_2018}
Sean Yiu and Li~Su.
\newblock Covariate association eliminating weights: a unified weighting
  framework for causal effect estimation.
\newblock \emph{Biometrika}, 105\penalty0 (3):\penalty0 709--722, September
  2018.
\newblock ISSN 0006-3444, 1464-3510.
\newblock \doi{10.1093/biomet/asy015}.
\newblock URL \url{https://academic.oup.com/biomet/article/105/3/709/4986429}.

\bibitem[Tübbicke(2022)]{tubbicke_entropy_2022}
Stefan Tübbicke.
\newblock Entropy {Balancing} for {Continuous} {Treatments}.
\newblock \emph{Journal of Econometric Methods}, 11\penalty0 (1):\penalty0
  71--89, January 2022.
\newblock ISSN 2156-6674.
\newblock \doi{10.1515/jem-2021-0002}.
\newblock URL
  \url{https://www.degruyterbrill.com/document/doi/10.1515/jem-2021-0002/html}.

\bibitem[Chipman et~al.(2010)Chipman, George, and McCulloch]{chipman_bart_2010}
Hugh~A. Chipman, Edward~I. George, and Robert~E. McCulloch.
\newblock {BART}: {Bayesian} additive regression trees.
\newblock \emph{The Annals of Applied Statistics}, 4\penalty0 (1), March 2010.
\newblock ISSN 1932-6157.
\newblock \doi{10.1214/09-AOAS285}.
\newblock URL
  \url{https://projecteuclid.org/journals/annals-of-applied-statistics/volume-4/issue-1/BART-Bayesian-additive-regression-trees/10.1214/09-AOAS285.full}.

\bibitem[Kreif et~al.(2015)Kreif, Grieve, Díaz, and
  Harrison]{kreif_evaluation_2015}
Noémi Kreif, Richard Grieve, Iván Díaz, and David Harrison.
\newblock Evaluation of the {Effect} of a {Continuous} {Treatment}: {A}
  {Machine} {Learning} {Approach} with an {Application} to {Treatment} for
  {Traumatic} {Brain} {Injury}.
\newblock \emph{Health Economics}, 24\penalty0 (9):\penalty0 1213--1228,
  September 2015.
\newblock ISSN 1057-9230, 1099-1050.
\newblock \doi{10.1002/hec.3189}.
\newblock URL \url{https://onlinelibrary.wiley.com/doi/10.1002/hec.3189}.

\bibitem[L'Ecuyer(1996)]{lecuyer_combined_1996}
Pierre L'Ecuyer.
\newblock Combined {Multiple} {Recursive} {Random} {Number} {Generators}.
\newblock \emph{Operations Research}, 44\penalty0 (5):\penalty0 816--822,
  October 1996.
\newblock ISSN 0030-364X.
\newblock \doi{10.1287/opre.44.5.816}.
\newblock URL \url{https://pubsonline.informs.org/doi/10.1287/opre.44.5.816}.

\bibitem[Fleming and Wallace(1986)]{fleming_how_1986}
Philip~J. Fleming and John~J. Wallace.
\newblock How not to lie with statistics: the correct way to summarize
  benchmark results.
\newblock \emph{Communications of the ACM}, 29\penalty0 (3):\penalty0 218--221,
  March 1986.
\newblock ISSN 0001-0782, 1557-7317.
\newblock \doi{10.1145/5666.5673}.
\newblock URL \url{https://dl.acm.org/doi/10.1145/5666.5673}.

\end{thebibliography}

\end{document}